\documentstyle[seceq,preprint,mbf]{ptptex}

\def\slash#1{\ooalign{\hfil/\hfil\crcr$#1$}}

\def\XZ{{\mbf Z}}
\def\XP{{\mbf P}}
\def\XQ{{\mbf Q}}
\def\XM{{\mbf M}}

\def\XD{{\mbf D}}
\def\XS{{\mbf S}}
\def\XK{{\mbf K}}
\def\XU{{\mbf U}}
\def\XX{{\mbf X}}
\def\XG{{\mbf G}}

\def\dt{\!\cdot\!}
\def\nn{\nonumber\\}

\def\calL{{\cal L}}
\def\calA{{\cal A}}
\def\calD{{\cal D}}
\def\hatD{{\hat\calD}}

\def\slashD{{\hat{\slash\calD}}}

\def\hatG{{\hat G}}
\def\hatR{{\hat R}}
\def\calF{{\cal F}}

\def\calR{{\cal R}}
\def\half{\hbox{\large ${1\over2}$}}
\def\myfrac#1#2{\hbox{\large${#1\over#2}$}}
\def\T{{\rm T}}
\def\c{{\rm c}}
\def\s{{\rm s}}

\def\6#1{{\underline{#1}}}
\def\m6#1{{\underline{#1}\,}}
\def\hA{{\cal A}}  \def\hF{{\cal F}}
\def\gA{A} \def\gB{B} \def\gC{C} 
\newdimen\Tdim
\def\ispan{{\setbox0=\hbox{i}%
\Tdim\ht0\advance\Tdim\dp0\rule[-\dp0]{0pt}{\Tdim}}}
\def\jspan{{\setbox0=\hbox{j}%
\Tdim\ht0\advance\Tdim\dp0\rule[-\dp0]{0pt}{\Tdim}}}
\def\Tspan#1{{\setbox0=\hbox{#1}%
\Tdim\ht0\advance\Tdim\dp0\advance\Tdim.55ex\rule[-\dp0]{0pt}{\Tdim}\box0}}

\pubinfo{Vol. 104, No. 4, October 2000}  
\notypesetlogo  
\preprintnumber[3cm]{
KUNS-1672\\ hep-ph/0006231}

\title{
Supergravity Tensor Calculus in 5D from 6D
}

\author{
Taichiro {\sc Kugo}\footnote{E-mail:
kugo@gauge.scphys.kyoto-u.ac.jp}
and Keisuke {\sc Ohashi}\footnote{E-mail:
keisuke@gauge.scphys.kyoto-u.ac.jp}
}

\inst{
Department of Physics, Kyoto University, Kyoto 606-8502, Japan
}

\recdate{
June 26, 2000}

\abst{
Supergravity tensor calculus in five spacetime dimensions is derived 
by dimensional reduction from the $d=6$ superconformal tensor calculus. 
In particular, we obtain an off-shell hypermultiplet in 5D from the 
on-shell hypermultiplet in 6D. Our tensor calculus retains the 
dilatation gauge symmetry, so that it is a trivial gauge fixing to 
make the Einstein term canonical in a general 
matter-Yang-Mills-supergravity coupled system.}

\begin{document}

\maketitle

\section{Introduction}

Many physicists have recently been studying seriously the revolutionary 
idea that our four-dimensional world may be a `3-brane' embedded in a 
higher dimensional spacetime. This idea should provide us with many 
new ideas and demands reconsideration of various problems in unification 
theories; e.g., gauge hierarchy,\cite{hierarchy} 
supersymmetry breaking,\cite{susy,ref:MP} hidden sectors,\cite{LutySund} 
fermion mass hierarchy,\cite{fermion} 
cosmology and astrophysics.\cite{astro}

This idea originally came from string theory,\cite{ref:Pol} 
which now has become recognized as not merely the theory of strings, but the
theory describing the totality of various dimensional branes. Ho\v rava 
and Witten\cite{ref:HW} introduced this type of idea for the first 
time in their investigation of the strongly-coupled heterotic string 
theory. They argued that the low-energy limit is described by 
eleven-dimensional supergravity compactified on $S^1/Z_2$, an interval 
bounded by orientifold planes, and that a ten-dimensional $E_8$ super 
Yang-Mills theory appears on each plane. Regarding one of the planes as 
`our world' and the other as `hidden', they offered an interesting proposal
for resolving the discrepancy between the GUT scale and the gravity 
scale.

As a toy model for more realistic phenomenology, in which 6 of the 
transverse 10 dimensions should be compactified, 
Mirabelli and Peskin\cite{ref:MP} considered a five-dimensional super 
Yang-Mills theory compactified on $S^1/Z_2$ and clarified how 
supersymmetry breaking occurring on one boundary is communicated to 
another. They presented a simple algorithm for coupling the bulk super 
Yang-Mills theory to the boundary fields with the help of off-shell 
formulations. It is clearly important to generalize their work to 
more realistic models in which the bulk is described by five-dimensional 
supergravity. 

One (presumably, the main) obstacle to this investigation has been the 
lack of an off-shell formulation of five dimensional supergravity. It 
is quite recent that such an off-shell formulation was first given by 
Zucker.\cite{ref:Zucker} \ In a series of two papers, he presented an 
almost complete supergravity tensor calculus, including the 40+40 minimal
Weyl multiplet, Yang-Mills multiplet, linear and nonlinear multiplets, 
and an invariant action formula applicable to the linear multiplet. 
Unfortunately and strangely, however, the most important matter 
multiplet in five dimensions, called a hypermultiplet or scalar multiplet,
was missing there. A general system of five-dimensional supergravity 
should necessarily contain hypermultiplets. Actually, an explicit form 
of the action for the general matter-Yang-Mills system coupled to 
supergravity would be very important and useful for phenomenological 
work, as we know from the example of the work of Cremmer et 
al.\cite{ref:CFGVP} on $N=1$, $d=4$ supergravity theory.

In this paper, we present a more complete 
tensor calculus for 5D supergravity including, in particular, the 
hypermultiplet. We derive it using dimensional reduction from the known 
$d=6$ superconformal tensor calculus, which was fully described in an 
excellent and elaborate paper by Bergshoeff, Sezgin and 
Van Proeyen,\cite{ref:BSVP} referred to as BSVP henceforth. 
The advantages of this dimensional reduction are two fold: One is 
obviously that we need not repeat the trial-and-error method to find the 
multiplet members and their transformation laws. Everything is in 
principle straightforward, and we can use all the formulas known in 6D. 
Secondly, the supergravity structure, like supersymmetry transformation 
laws, is actually simpler in 6D than in 5D, because of its larger symmetry. 
Knowing the relations between covariant derivatives and curvatures in the 
5D and 6D theories, we can obtain deeper understanding of the group 
structure of 5D supergravity. (Even in applications other than 
constructing 5D tensor calculus, such knowledge may become useful.) 
Moreover, we can inherit the advantage of superconformal symmetry of the 
original 6D theory. Indeed, we retain the dilatation gauge symmetry also
in our 5D tensor calculus, and this makes it so that a trivial 
gauge-fixing renders the Einstein-Rarita-Schwinger term 
canonical.\cite{ref:KU2} 

The rest of this paper is organized as follows. 
In \S2, we explain in some detail the dimensional reduction procedure 
to obtain our 5D supergravity tensor calculus from the 6D 
superconformal one of BSVP. We also present there some general discussion 
on the covariant derivatives and curvatures, which applies both to 
supergravity and superconformal theories and turns out to be very useful.
Based on this, the transformation law is derived for the 40+40 minimal 
Weyl multiplet in 5D. Then the transformation laws of Yang-Mills, 
linear and nonlinear multiplets can be found straightforwardly, as given 
in \S3. However, we need a method to obtain an {\it off-shell} 
hypermultiplet in 5D from the 6D one which exists only as an on-shell 
multiplet. As explained in \S4, we make it off-shell in our reduction 
procedure using a method similar to that known in $d=4$ 
case.\cite{ref:onoff} The invariant action formulas for the kinetic 
terms as well as mass terms for such off-shell hypermultiplets are also 
derived there. Other invariant action formulas and some embedding 
formulas are given in 
\S5. The final section is devoted to summary and discussion.

We here note that we adopt the metric $\eta 
_{ab}=\hbox{diag}(+,-,-,\cdots,-)$, which is different from that of BSVP, 
but we think it is more familiar to phenomenologists. Our notation and 
conventions are given in Appendix A, where some useful formulas are also
given. The results of the 6D superconformal tensor calculus of BSVP that
we need in this paper are briefly summarized in Appendix B.

\section{Dimensional reduction and 5D Weyl multiplet}

\subsection{Dimensional reduction and gauge-fixing}

Our starting point is the superconformal tensor calculus in $d=6$ 
dimensions given by BSVP\cite{ref:BSVP} and summarized in Appendix B.
The superconformal group in six dimensions is 
$OSp(6,2|1)\simeq OSp(4|1;\mbf H)$ which has the generators 
\begin{equation}
\XP_{\6a},\ \XM_{\6a\6b},\ \XU_{ij},\ \XD,\ \ \XK_{\6a},\ 
\XQ_{\alpha i},\ \XS_{\alpha i},
\label{eq:6Dgener}
\end{equation}
where $\6a,\6b,\cdots$ are tangent vector indices, $\alpha$ is a spinor index, and
$i,j,\cdots=1,2$ are {\it SU}$(2)$ indices. 
$\XP_{\6a}$ and $\XM_{\6a\6b}$ represent the usual Poincar\'e generators, 
$\XU_{ij}$ represents the {\it SU}$(2)$ generators, $\XD$ is 
the dilatation, $\XK_{\6a}$ represents the special conformal boosts.  
$\XQ_{\alpha i}$ represents the supersymmetry and $\XS_{\alpha i}$ represents the 
conformal supersymmetry, both of which are {\it 
SU}$(2)$-Majorana-Weyl spinors. The gauge fields corresponding to these 
generators are
\begin{equation}
\m6e^{\6a}_{\6\mu},\ \m6\omega_{\6\mu}{}^{\6a\6b},\ \m6V_{\6\mu}^{\,ij},\ 
\m6b_{\6\mu},\ \m6f_{\6\mu}^{\6a},\ \m6\psi_{\6\mu}^i,\ \m6\phi_{\6\mu}^i,
\end{equation}
respectively, where the spinor indices of the 
gauge fields $\m6\psi_{\6\mu}^i$ and $\m6\phi_{\6\mu}^i$ are suppressed. 

The local superconformal algebra in $d=6$ can only be realized by adding a 
suitable `matter' multiplet to these gauge fields, and 
the $\XM$, $\XK$ and $\XS$ gauge fields, 
$\m6\omega_{\6\mu}{}^{\6a\6b},\ \m6f_{\6\mu}^{\6a},\ \m6\phi_{\6\mu}^i$, become 
dependent fields through a set of constraints, (\ref{eq:6D.cons}), 
imposed on the curvatures. Then, the algebra is realized on the 
set of 40 Bose plus 40 Fermi fields 
\begin{equation}
\m6e^{\6a}_{\6\mu},\ \m6V_{\6\mu}^{\,ij},\ 
\m6b_{\6\mu},\ \m6\psi_{\6\mu}^i,\ T^-_{\6a\6b\6c},\ \chi^i,\ D,
\label{eq:WeylM}
\end{equation}
called minimal {\it Weyl multiplet}, where the last three fields,
an anti-self-dual tensor 
$T^-_{\6a\6b\6c}$,  an {\it SU}$(2)$-Majorana-Weyl spinor $\chi^i$, and  
a real scalar $D$, are the added `matter multiplet'.

We generally add underlines to the quantities in 6 dimensions when 
necessary to distinguish them from those in 5 dimensions. In particular,
the Lorentz index is $\6a=(a,5)$, and the world vector index is 
${\6\mu}=(\mu,z)$; we use $z$ to denote both the fifth spatial direction 
and the coordinate itself, $x^{\6\mu}=(x^\mu,x^z=z)$, since 
we must distinguish the curved index fifth spatial component 
$v_z\equiv v_{\6\mu=5}$ from the flat index one 
$v_5\equiv v_{\6a=5}$. 

We now perform a `trivial' dimensional reduction from 6 to 5 dimensions, 
by simply letting all the fields and local transformation parameters 
be independent of the fifth spatial coordinate $z$. As in the usual 
procedure,\cite{ref:CJS} we fix the off-diagonal local Lorentz 
$\XM_{a5}$ gauge by setting $\m6e_z^{\,a}=0$, and then the sechsbein 
$\m6e_{\6\mu}^{\6a}$ and its inverse take the forms
\begin{equation}
\m6e_{\6\mu}^{\,\6a}
=\pmatrix{
e_\mu^{\,a} &  e_\mu^{\,5}=\alpha^{-1}A_\mu \cr
e_z^{\,a}=0       &  e_z^{\,5}=\alpha^{-1}  \cr}, \qquad 
\m6e_{\6a}^{\,\6\mu}=\pmatrix{
e_a^{\,\mu} & e_a^{\,z}=-e_a^{\,\mu}A_\mu\cr
e_5^{\,\mu}=0       &  e_5^{\,z}=\alpha\cr}.
\label{eq:sechsbein}
\end{equation}

As is well-known since the work of Kaluza and Klein, the field $A_\mu$ 
appearing in the off-diagonal element $e_\mu^{\,5}$ becomes a $U(1)$ 
gauge field which we call a `gravi-photon'. Under the general coordinate 
(GC) transformation $\6\delta_{\rm GC}(\xi^{\6\nu})$ in 6D with the 
transformation parameter $\xi^{\6\mu}=(\xi^\mu,\xi^z)$ taken to be 
$z$-independent, any field with a world vector index (e.g., gauge 
fields) $\6h_{\6\mu}=(\m6h_\mu,\m6h_z)$ transforms as
\begin{eqnarray}
\6\delta_{\rm GC}(\xi^{\6\nu})\m6h_\mu&=&
\partial_\mu\xi^{\6\nu}\cdot \m6h_{\6\nu} + \xi^\nu\partial_\nu\m6h_\mu 
=\delta_{\rm GC}(\xi^{\nu})\m6h_\mu+\partial_\mu\xi^{z}\cdot \m6h_z\,,
\label{eq:6DGC} \\
\6\delta_{\rm GC}(\xi^{\6\nu})\m6h_z&=&
\xi^\nu\partial_\nu\m6h_z=\delta_{\rm GC}(\xi^{\nu})\m6h_z\,, 
\end{eqnarray} 
where $\delta_{\rm GC}(\xi^{\nu})$ is the GC transformation in 5D. 
For the gravi-photon $A_\mu=e_\mu^{\,5}e_5^{\,z}$,
in particular, we have 
\begin{equation}
\6\delta_{\rm GC}(\xi^{\6\nu})A_\mu=
\delta_{\rm GC}(\xi^{\nu})A_\mu+\partial_\mu\xi^{z},
\label{eq:ZonA}
\end{equation} 
showing that $A_\mu$ actually transforms as the $U(1)$ gauge field 
for the transformation parameter $\xi^z$. The discrepancy between the 
6D and 5D GC transformations on the other vector fields $\m6h_\mu$ 
[the second term $\partial_\mu\xi^z\cdot\m6h_z$ in Eq.~(\ref{eq:6DGC})] can thus 
be eliminated by redefining the field as
\begin{equation}
h_\mu\equiv\m6h_\mu-A_\mu\m6h_z,
\end{equation}
which we, therefore, identify with the vector (or gauge) field in 5D. 
But we note that this identification rule can be rephrased into a 
simpler rule:\cite{ref:CJS} 
{\em Any field with flat (local Lorentz) indices alone is the same 
in 6D and 5D}. \,Indeed, if converted into flat indices, we have 
$\m6h_a = e_a^\mu\m6h_\mu+e_\mu^z\m6h_z
=e_a^\mu(\m6h_\mu-A_\mu\m6h_z)=e_a^\mu h_\mu=h_a$. We hence adopt this rule 
throughout in our dimensional reduction. Thus we can omit the 
underlines for the fields with flat indices alone. The $z$-component 
$\m6h_z$ is just a scalar in 5D and may be denoted $h_z$ without an 
underline. For the fields with upper curved vector indices, 
we have $\m6h^\mu=h^\mu$, $\m6h^z=\alpha h^5-A_ah^a$. 

Now, the 6D GC transformation $\6\delta_{\rm GC}(\xi^{\6\nu})$, which appears 
in the supersymmetry transformation commutator 
$[\m6\delta_Q(\varepsilon_1),\m6\delta_Q(\varepsilon_2)]$ 
with $\xi^{\6\nu}=2i\bar\varepsilon_1\gamma^{\6\nu}\varepsilon_2$, 
reduces to the 5D one, $\delta_{\rm GC}(\xi^{\nu})$, {\it plus} the $U(1)$ 
gauge transformation $\delta_Z(\xi^z)$, which acts only on the gravi-photon 
field $A_\mu$ among the gauge fields as
\begin{equation}
\delta_Z(\theta)A_\mu=\partial_\mu\theta.
\label{eq:Zdef}
\end{equation}
Below, we identify this transformation $\delta_Z(\theta)$ with the central 
charge transformation acting on the hypermultiplet fields, which is 
originally the fifth spatial derivative $\partial_z$ in their supersymmetry 
transformation law in 6D.

The spinor fields in 6D superconformal theory are all 
{\it SU}$(2)$-Majorana-Weyl spinors $\psi_\pm^i$ 
satisfying simultaneously the {\it SU}$(2)$-Majorana condition
\begin{equation}
\bar\psi_\pm^i \equiv(\psi_{\pm i})^\dagger\6\gamma^0 = (\psi_\pm^i)^{\rm T}\6C\,,
\end{equation}
and the Weyl condition (of positive or negative chirality)
\begin{equation}
\6\gamma_7\psi_\pm^i = \pm\psi_\pm^i,
\end{equation}
where $\6\gamma^{\6a}$ and $\6C$ are $8\times8$ Dirac gamma and charge 
conjugation matrices in 6D, and $\6\gamma_7\equiv-\6\gamma^0\6\gamma^1\cdots\6\gamma^5$. 
To make contact with the 4-component spinors in 5D, we can use the 
following representation of the 6D gamma matrices $\6\gamma^{\6a}$ 
given in terms of the 5D $4\times4$ ones $\gamma^a$ satisfying 
$\gamma^0\gamma^1\gamma^2\gamma^3\gamma^4=1$:
\begin{equation}
\cases{
\6\gamma^a =\gamma^a\otimes \sigma_1      \qquad \hbox{for}\  a=0,1,2,3,4\,, \cr
\6\gamma^5 ={\bf 1}_4\otimes i\sigma_2\,, \cr
\6\gamma_7 =-\6\gamma^0\6\gamma^1\cdots\6\gamma^5= {\bf 1}_4 \otimes \sigma_3\,. \cr}
\end{equation}
The charge conjugation matrix $\6C$ in 6D is given by the 
5D one $C$ as
\begin{equation}
\6C=C \otimes i\sigma_2\,.
\end{equation}
Then the {\it SU}$(2)$-Majorana-Weyl spinors $\psi_\pm^i$ in 6D reduce to the forms 
\begin{equation}
\psi^i_+=
\pmatrix{\psi^i \cr 0 \cr}, \qquad 
\psi^i_-=
\pmatrix{0 \cr i\psi^i \cr},
\label{eq:spinorREL}
\end{equation}
and both $\psi^i$ are now 
the 4-component {\it SU}$(2)$-Majorana spinors in 5D satisfying
\begin{equation}
\bar\psi^i \equiv(\psi_i)^\dagger\gamma^0 = (\psi^i)^{\rm T}C\,.
\end{equation}
Thus the spinors appearing in 5D are all of the {\it SU}$(2)$-Majorana 
type, and we generally use the same symbols to denote the spinors in 5D as 
those used by BSVP, although they are actually related by 
Eq.~(\ref{eq:spinorREL}).

In reducing to 5 dimensions by setting $e_z^{\,a}=0$, it turns out to
be simpler to also fix the conformal $\XS$ and $\XK$ gauge symmetries 
by using the following gauge-fixing conditions:
\begin{eqnarray}
&&\XM_{a5}:\ \ e_z^{\,a}=0, \qquad \XS^i:\ \ \psi^i_5=0, \nn
&&\XK_{a}:\ \ b_\mu-\alpha^{-1}\partial_\mu\alpha=0, \qquad \XK_5:\ \ b_5=0.
\label{eq:GaugeCond}
\end{eqnarray}
The $\XS$ gauge is chosen to satisfy $\psi^i_5=0$ (implying also $\psi 
_z=e_z^{\,5}\psi_5=0$), so as to make the condition $e_z^{\,a}=0$ 
invariant under the supersymmetry transformation, $\m6\delta_Q(\varepsilon)
e_z^{\,a}=-2i\bar\varepsilon\6\gamma^a\psi_z$. Note that this gauge also implies the 
$\XQ$-invariance of the `dilaton' field $\alpha\equiv 
(e_z^{\,5})^{-1}=e_5^{\,z}$. Moreover, the $\XK_a$ gauge $b_\mu=\alpha^{-1}\partial 
_\mu\alpha$ is chosen so as to make $\alpha$ also covariantly constant $\hatD_\mu 
\alpha=\partial_\mu\alpha-b_\mu\alpha=0$ in the reduced 5D theory, as we see below. 
Thanks to these two properties, the field $\alpha$ carrying Weyl weight 
$w=1$ can be treated as if it were a constant and is freely used to 
adjust the Weyl weights of any quantities to arbitrary desired values. 
We, however, keep the dilatation gauge symmetry unfixed, since it 
becomes useful later when we change the Einstein-Hilbert and 
Rarita-Schwinger terms in the action into canonical form. We here 
note that the extraneous components $\m6\omega_{\mu}{}^{a5}$ and $\m6\omega 
_{5}{}^{\6a\6b}$ of the spin connection $\m6\omega_{\6\mu}{}^{\6a\6b}$ given 
by Eq.~(\ref{eq:6Dspinconn}) now take the following simple form under 
these gauge-fixing conditions:
\begin{equation}
e^{a\mu}\omega_\mu{}^{b5}=\omega_5{}^{ab}=
-\myfrac1{2\alpha}\hat F^{ab}(A), \qquad 
\omega_5{}^{a5}=0.
\end{equation}
Here $\hat F_{\mu\nu}(A)$ is the supercovariantized field strength of the 
gravi-photon:
\begin{equation}
\hat F_{\mu\nu}(A)=F_{\mu\nu}(A) -2i\alpha\bar\psi_\mu\psi_\nu, \qquad 
F_{\mu\nu}(A) 
= 2\partial_{[\mu}A_{\nu]}.
\end{equation}

Here a summary of the 5D field contents and their notation may be
in order. The fields we are now treating all come from the 40+40 Weyl 
multiplet (\ref{eq:WeylM}) in 6D, and we also call them a 5D Weyl multiplet.
Since $\psi_5$ and $b_{\6\mu}$ are eliminated by the above gauge fixing, 
the remaining fields are now
\begin{eqnarray}
&&\m6e^{\6a}_{\6\mu} \ \rightarrow\  \cases{e_\mu^a \cr A_\mu\cr \alpha}, \qquad  
\m6V_{\6\mu}^{\,ij} \ \rightarrow\ \cases{V_{\mu}^{\,ij} \cr t^{ij}\equiv-V_5^{ij} \cr},
\qquad 
\m6\psi_{\6\mu}^i \ \rightarrow\ \psi_\mu^i, \nn
&&T^-_{\6a\6b\6c} \ \rightarrow\  
v_{ab}\equiv-T^-_{ab5}+\myfrac1{4\alpha}\hat F_{ab}(A),\qquad 
\chi^i\rightarrow\chi^i,\qquad 
D \rightarrow D.
\label{eq:5DWeylM}
\end{eqnarray}
The quantities on the right-hand sides of the arrows here, 
$e_\mu^a,\  A_\mu,\ \alpha,\ V_{\mu}^{\,ij},\ t^{ij},\ 
\psi_\mu^i,\ v_{ab}$, $\chi_i$ and $D$, denote our 5D fields.
\begin{table}[tb]
\caption{Weyl multiplet in 5D.}
\label{table:5DWeyl}
\begin{center}
\begin{tabular}{ccccc} \hline \hline
    field      & type   & restrictions & {\it SU}$(2)$ & Weyl-weight    \\ \hline 
\Tspan{$e_\mu{}^a$} &   boson    & f\"unfbein    & \bf{1}    &  $ -1$     \\  
$\psi^i_\mu$  &  fermion  & {\it SU}$(2)$-Majorana & \bf{2}  &$-\hspace{-1.5mm}\myfrac12$ \\  
\Tspan{$V^{ij}_\mu$}    &  boson    & $V_\mu^{ij}=V_\mu^{ji}=-V^*_{\mu ij}$ 
& \bf{3}&0\\ 
$A_\mu$    &  boson    & gravi-photon $A_\mu=e^z_5e_\mu^5$   & \bf{1}&0 \\ 
$\alpha$    &  boson    & `dilaton' $\alpha=e_5^z$   & \bf{1}& 1 \\ 
$t^{ij}$    &  boson    & $t^{ij}=t^{ji}=-t^*_{ij}\ (=-V_5^{ij})$   & \bf{3} & 1 \\ 
$v_{ab}$&boson&$v_{ab}=-T^-_{ab5}+\myfrac1{4\alpha}\hat{F}_{ab}(A)$&\bf{1}&1 \\
$\tilde \chi^i$  &  fermion  & {\it SU}$(2)$-Majorana & \bf{2}    &\hspace{-1.1mm}\myfrac32 \\  
$C$    &  boson    & real & \bf{1} & 2 \\ \hline
\multicolumn{5}{c}{dependent gauge fields} \\ \hline
\Tspan{$b_\mu$} & boson &  $\XD$ gauge field $b_\mu=\alpha^{-1}\partial_\mu\alpha$ 
& \bf{1} & 0 \\
$\omega_\mu{}^{ab}$ &   boson    & spin connection & \bf{1}    &   0     \\ 
\hline 
\end{tabular}
\end{center}
\end{table}
The fields $t^{ij}$ and $v_{ab}$ are defined to be particular 
combinations of the fields in order to simplify the expressions of the 
supersymmetry transformation in 5D. In the same sense, it turns out to 
be convenient to use the following spinor field $\tilde\chi^i$ 
and scalar field $C$ in place of $\chi^i$ and $D$, respectively:
\begin{eqnarray}
\tilde\chi^i&\equiv&
\myfrac1{15}(\chi^i+\myfrac34\gamma^{ab}\hat{\cal R}_{ab}{}^i(Q))
=\myfrac1{12}\chi^i+2\phi_5^i\,,\nn
C&\equiv&\myfrac1{15}\left(D-\myfrac34\hat{\cal R}(M)+15v\dt v 
-\myfrac9{8\alpha^2}\hat F(A)\dt \hat F(A)-30t^i{}_jt^j{}_i\right)\,.
\label{eq:deflambda}
\end{eqnarray}
Here $\hat{\calR}_{ab}{}^i(Q)$ and $\hat{\cal R}(M)$ are the curvatures 
in 5D theory discussed in detail below, and $\phi_5$ is the fifth spatial 
component of the 6D $\XS$-gauge field $\phi_{\6a}$. The members of the 5D 
Weyl multiplet are listed in Table \ref{table:5DWeyl}.

The supersymmetry transformation in 5D is found as follows. 
The $\XS$ and $\XK_{\6a}$ gauge-fixing conditions in (\ref{eq:GaugeCond}) 
are not invariant under the original $\XQ$ transformation 
$\m6\delta_Q(\varepsilon)$ in 6D. The deviations of 
$\m6\delta_Q(\varepsilon)\psi^i_5$ and $\m6\delta_Q(\varepsilon)(b_\mu-\alpha^{-1}\partial_\mu\alpha)$ from zero 
of course can be transformed back to zero by suitable 
$\XS$ and $\XK_{\6a}$ gauge transformations. 
Thus the following combined transformation of $\XQ$, $\XS$ and $\XK$ is
found to leave these conditions intact, and can be defined to give 
the supersymmetry transformation $\delta_Q(\varepsilon)$ in 5D:
\begin{eqnarray}
&&\hspace{-6.3em}
\delta_Q(\varepsilon)=\m6\delta_Q(\varepsilon)+\delta_S(\eta(\varepsilon))+\delta_K(\xi^{\6a}_K(\varepsilon))\,,
\label{eq:SusyTrf} \nn
\hbox{with}\quad  
\eta(\varepsilon)^i
&=&-\myfrac1{16\alpha}\gamma\dt \hat F(A)\varepsilon^i-\myfrac14\gamma\dt v\varepsilon^i-t^i_j\varepsilon^j\,, \nn
\xi_K^a(\varepsilon)&=&-i\bar\varepsilon(\phi^a-\eta(\psi^a)+\myfrac{1}{24}\gamma^a\chi)\,, \nn
\xi_K^5(\varepsilon)&=& 
i\bar\varepsilon(\phi_5+\myfrac1{24}\chi) =\half i\bar\varepsilon\tilde\chi\,.
\end{eqnarray}
Here, $\gamma\dt T$ for any tensor $T_{a_1\cdots a_n}$ generally denotes the
contraction $\gamma^{a_1\cdots a_n}T_{a_1\cdots a_n}$. Note that the spinors in these 
expressions and the following already stand for the 5D 4-component 
spinors defined on the right-hand side of Eq.~(\ref{eq:spinorREL}). 

Now it is straightforward to obtain the supersymmetry transformation laws 
of the Weyl multiplet in 5D from the superconformal transformation laws 
(\ref{eq:6DWeylTrf}) in 6D by using Eq.~(\ref{eq:SusyTrf}). Before doing
this, however, it is better to define the covariant derivatives and to 
derive the relations between curvatures in 6D and 5D, since they appear 
in the supersymmetry transformation laws. 

\subsection{Covariant derivative and curvatures}

We now give somewhat general discussion on the curvatures in supergravity. 
(Also see the discussion in Refs.~\citen{ref:Sohnius} and \citen{ref:VP}.) 
\ Let $\{\XX_{\bar A}\}$ denote a set of local transformation operators 
acting on the fields $\phi$, $\varepsilon^{\bar A}\XX_{\bar A}\phi=\delta_{\bar A}(\varepsilon)\phi 
$, and satisfy the graded algebra
\begin{equation}
[\XX_{\bar A},\,\XX_{\bar B}\}=f_{{\bar A}{\bar B}}{}^{\bar C}\XX_{\bar C}. 
\label{eq:Galg}
\end{equation}
$f_{{\bar A}{\bar B}}{}^{\bar C}$ here in general 
depends on the fields and we call it ``structure function''. 
In the (Poincar\'e or conformal) supergravity theory, the set 
$\{\XX_{\bar A}\}$ 
contains the `translation' $\XP_a$, which plays a special role.
The transformation operators other than $\XP_a$ are denoted 
$\XX_A$ with no bar over the index $A$. Let us now introduce two kinds of 
covariant derivatives, excluding and including $\XP_a$ covariantization:
\begin{equation}
\hatD_\mu\phi\equiv\partial_\mu\phi-h_\mu^{A}\XX_{A}\phi,\qquad 
{\mbf\nabla}_\mu\phi\equiv\partial_\mu\phi- h_\mu^{\bar A}\XX_{\bar A}\phi 
=\hatD_\mu\phi-e_\mu^a\XP_a\phi.
\label{eq:covD}
\end{equation}
Here, $h_\mu^{\bar A}$ are gauge fields, and sums over the repeated 
indices $A$ and $\bar A$ are implied. The operator ${\mbf\nabla}_\mu$ is the 
`usual' covariant derivative, fully covariant with respect to all the 
gauge transformations, while $\hatD_\mu$ is the covariant derivative 
adopted in supergravity. Then, as in Yang-Mills theory, imposing the 
covariance of the `usual' one ${\mbf\nabla}_\mu$ [i.e. $\XX_{\bar A}({\mbf
\nabla}_\mu\phi)={\mbf\nabla}_\mu(\XX_{\bar A}\phi)$] determines the transformation 
law of the gauge fields as
\begin{equation}
\varepsilon^{\bar A}\XX_{\bar A}h_\mu^{\bar A}\equiv\delta(\varepsilon)h_\mu^{\bar A} 
= \partial_\mu\varepsilon^{\bar A}+\varepsilon^{\bar C}h_\mu^{\bar B}f_{{\bar B}{\bar C}}{}^{\bar A},
\label{eq:gaugeTrf}
\end{equation}
and the commutator of ${\mbf\nabla}_\mu$ defines the curvature tensors 
${\mbf R}_{\mu\nu}{}^{\bar A}$ in the form
\begin{equation}
{}[{\mbf\nabla}_\mu,\,{\mbf\nabla}_\nu] = -{\mbf R}_{\mu\nu}{}^{\bar A}\XX_{\bar A}
\quad \rightarrow\quad {\mbf R}_{\mu\nu}{}^{\bar A}\equiv2\partial_{[\mu}h_{\nu]}^{\bar A}
-h_\mu^{\bar C}h_\nu^{\bar B}f_{\bar B \bar C}{}^{\bar A}.
\label{eq:RYMdef}
\end{equation}

Now, the particular feature of supergravity is the stipulation that 
the `usual' covariant derivative ${\mbf\nabla}_\mu$ vanish on any matter 
field $\phi$ carrying flat indices alone:
\begin{equation}
{\mib\nabla}_\mu\phi=0 \qquad \rightarrow\qquad \hatD_\mu\phi=e_\mu^a\XP_a\phi.
\label{eq:SGdemand}
\end{equation}
In supergravity, thus, the only meaningful covariant derivative is
$\hatD_\mu$, whose flat index version, $\hatD_a=e_a^{\,\mu}\hatD_\mu$, 
gives meaning to the `translation' transformation $\XP_a$.
This stipulation embodied by (\ref{eq:SGdemand}) can be imposed if and 
only if
\begin{equation}
{}[{\mbf\nabla}_\mu,\,{\mbf\nabla}_\nu] = 0 
\quad \rightarrow\quad {\mbf R}_{\mu\nu}{}^{\bar A}=0
\end{equation}
is satisfied. The curvature in supergravity, $\hatR_{ab}{}^{\bar A}$, 
is defined via the flat $\hatD_a$ by 
\begin{equation}
{}[\hatD_a,\,\hatD_b] 
= -\hatR_{ab}{}^{\bar A}\XX_{\bar A}
\equiv-\hatR_{ab}.
\label{eq:Rdef}
\end{equation}
Without carrying out cumbersome computations going back to the original 
definition of $\hatD_a$, we can immediately find the following simple 
expression for this curvature,
\begin{equation}
\hatR_{\mu\nu}{}^{\bar A}
= e_\mu^{\,b}e_\nu^{\,a}f_{ab}{}^{\bar A}
=2\partial_{[\mu}h_{\nu]}^{\bar A}
-h_\mu^{\bar C}h_\nu^{\bar B}f'_{\bar B \bar C}{}^{\bar A},
\label{eq:Rexpr}
\end{equation} 
where the prime on the structure function indicates that 
the $[\XP_a,\,\XP_b]$ commutator parts $f_{ab}{}^{\bar A}$ 
are excluded from the sum. The first equality follows from the 
relation $\hatD_a=\XP_a$, holding on any flat-indexed fields $\phi$, and
\begin{equation}
-\hatR_{ab}{}^{\bar A}
=[\hatD_a,\hatD_b]^{\bar A}=
[\XP_a,\XP_b]^{\bar A}=
f_{ab}{}^{\bar A},
\end{equation}
and the second equality follows from ${\mbf R}_{\mu\nu}{}^{\bar A}=0$ and 
Eq.~(\ref{eq:RYMdef}). Conversely, if $\hatR_{\mu\nu}{}^{\bar A}$ is 
defined by Eq.~(\ref{eq:Rexpr}), then Eq.~(\ref{eq:Rdef}) follows, of 
course, and it can be used as a convenient formula. Another convenient 
formula also follows immediately from 
$[\XX_A,\hatD_a]=[\XX_A,\XP_a]=f_{Aa}{}^{\bar B}\XX_{\bar B}$ for the 
transformation $\delta(\varepsilon)\equiv\varepsilon^A\XX_A$ not including $\XP_a$:
\begin{equation}
\delta(\varepsilon)\hatD_a\phi 
= \varepsilon^A \hatD_a(\XX_A\phi) + \varepsilon^Af_{Aa}{}^{\bar B}\XX_{\bar B}\phi.
\end{equation}

Using the Jacobi identity 
$[\XX,\,[\hatD_a,\,\hatD_b]\,]
+\hbox{(permutations)}=0$ and the additional 
information that $f_{ab}{}^c=-\hatR_{ab}{}^c(P)=0$ and 
$f_{aB}{}^c=\hbox{const}$, holding generally in supergravity, 
we can obtain the 
following Bianchi identities: For $\XX=\XP_c$, we have
\begin{eqnarray}
\hatD_{[a}\hat R_{bc]}{}^A&=&-\hat R_{[ab}{}^Bf_{c]B}{}^A, 
\label{eq:Bianchi}\\
\hat R_{[ab}{}^Af_{c]A}{}^d&=&0, 
\end{eqnarray}
and for $\XX=\XX_{A}$ and $\delta(\varepsilon)\equiv\varepsilon^A\XX_A$, we obtain
\begin{eqnarray}
\delta(\varepsilon)\hat R_{ab}{}^A &=&2\varepsilon^B\hatD_{[a}f_{b]B}{}^A
 +2\varepsilon^Bf_{B[a}{}^{\bar C}f_{b]\bar C}{}^A-\hat R_{ab}{}^C\varepsilon^Bf_{BC}{}^A, 
\label{eq:Rtrf}
\nn
\hat R_{ab}{}^B\varepsilon^Af_{AB}{}^c &=& 
2\varepsilon^Af_{A[a}{}^{\bar B}f_{b]\bar B}{}^c.
\end{eqnarray}
%
Finally in this general discussion, we add a remark on the meaning of 
the $\XP$ transformation $\delta_P(\xi)=\xi^a\XP_a$ on the gauge fields 
$h_\mu^{\bar A}$. The GC transformation of $h_\mu^{\bar A}$ 
given in Eq.~(\ref{eq:6DGC}) can be rewritten by using 
${\mbf R}_{\mu\nu}{}^{\bar A}$ in the form
\begin{equation}
\delta_{\rm GC}(\xi^\nu)h_\mu^{\bar A} = \xi^\nu h_\nu^{\bar B}\XX_{\bar B}h_\mu^{\bar A}
+\xi^\nu{\mbf R}_{\mu\nu}{}^{\bar A}.
\end{equation}
Using ${\mbf R}_{\mu\nu}{}^{\bar A}=0$ and extracting the $\XP_a$ term from 
$\XX_{\bar B}$, we find
\begin{equation}
\delta_P(\xi)h_\mu^{\bar A}= \left[\delta_{\rm GC}(\xi^\nu=e^\nu_a\xi^a) 
- \xi^ah_a^B\XX_{B}\right]h_\mu^{\bar A}.
\end{equation}
Comparing this with $\delta_P(\xi)\phi=\xi^a\hatD_a\phi 
=(\xi^\mu\partial_\mu-\xi^ah_a^B\XX_{B})\phi$ for the flat quantity 
$\phi$, the simple derivative term $\xi^\mu\partial_\mu$ is replaced by the GC 
transformation $\delta_{\rm GC}(\xi^\mu)$ here. Thus the replacement 
$\xi^\mu\partial_\mu\rightarrow\delta_{\rm GC}(\xi^\mu)$ 
should be generally understood in $\xi^a\hatD_a$ if acting on 
quantities with curved indices. This is a general rule, because the 
vielbein $e_\mu^a$ obeys it, and the conversion of flat indices into curved
ones is performed using the vielbein.

The discussion up to this point is general and applies, in particular, 
both to the present 6D superconformal theory and 5D supergravity, 
which we obtain from it. Again, to distinguish them, we 
write the covariant derivative and the curvatures in 6D with underlines 
as $\m6\hatD_a$ and $\m6\hatR_{\6a\6b}{}^A$, while those in 5D as 
$\hatD_a$ and $\hat{\calR}_{ab}{}^A$. 

To find relations between $\m6\hatR_{\6a\6b}{}^A$ and 
$\hat{\calR}_{ab}{}^A$ by using the formula (\ref{eq:Rdef}),
we first need the relation between the 
covariant derivatives $\m6\hatD_a$ and $\hatD_a$. 
Note that, in the 5D reduced theory, the transformations $\{ \XX_{\bar 
A} \}$ are only
\begin{equation}
\XP_{a},\ \XM_{ab},\ \XU_{ij},\ \XD,\ \XQ_{\alpha i},\,\XZ;
\label{eq:5Dgener}
\end{equation}
that is, $\XM_{a5},\,\ \XS_{\alpha i}$ and $\XK_{\6a}$ have disappeared 
in reducing from the 6D superconformal theory, whose generators are 
given in Eq.~(\ref{eq:6Dgener}). 
Using the definition (\ref{eq:covD}) of the covariant derivative and 
noting also the relation (\ref{eq:SusyTrf}) of the $\XQ$ transformations
in 5D and 6D, we find 
\begin{eqnarray}
{\6\hatD}_a&=&
 \hatD_a-\delta_{M_{b5}}(\omega_a{}^{b5})
             -\delta_S(\phi_a^i-\eta(\psi_a)^i)
             -\delta_{K_{\6b}}(f_a{}^{\6b}-\xi_K^{\6b}(\psi_a))\,,\nn
{\6\hatD}_5&=&\delta_Z(\alpha)-\delta_M(\omega_5{}^\6{ab})+\delta_U(t^{ij})
           -\delta_S(\phi_5^i)-\delta_{K_{\6a}}(f_5{}^{\6a}) -\delta_G(W_5)\,. 
\hspace{2em}
\label{eq:Drel}
\end{eqnarray}
Note here that the fifth spatial derivative $\partial_z$ in $\m6\hatD_{\6b}$ 
has been identified with the central charge transformation $\XZ$, so that 
${\6\hatD}_5$ contains $e_5^z\partial_z\rightarrow\alpha\XZ=\delta_Z(\alpha)$ and 
${\6\hatD}_a$ also contains $e_a^\mu\partial_\mu+e_a^z\partial_z=\partial_a-A_a\XZ$. 
The identification $\partial_z=\XZ$ holds exactly on the hypermultiplet, 
which we discuss below, and, moreover, it is also consistent with the
previous central charge term $\delta_Z(\xi^z)A_\mu=\partial_\mu\xi^z$, which appeared 
as a part of 6D GC transformation of the gravi-photon $A_\mu$ in 
Eq.~(\ref{eq:ZonA}). Actually, as remarked above, the derivative term $
\xi^z\partial_z$ in the covariant derivative $\xi^z\m6\hatD_z$ should be 
understood as the GC transformation $\m6\delta_{\rm GC}(\xi^z)$ on the 
gauge fields and, on $A_\mu$, it indeed yields 
$\m6\delta_{\rm GC}(\xi^z)A_\mu=\partial_\mu\xi^z$. 
This explains the reason that the gravi-photon is identified with the 
gauge field corresponding to the central charge of the hypermultiplet.
Note also that if Yang-Mills fields $\6W_{\6\mu}$ with gauge group $G$ 
are coupled to the system we should also include $G$-covariantization in
$\6\hatD_{\6a}$, so that $\hatD_a$ contains $-\delta_G(W_a)$ implicitly and 
$\6\hatD_5$ contains $-\delta_G(W_5)$ as written explicitly in the last term. 

Using the relations in (\ref{eq:Drel}) and the formula 
(\ref{eq:Rdef}) we obtain equations of the form
\begin{eqnarray}
\hat{\cal R}_{ab}&=&[\hatD_b,\,\hatD_a]
 =[\m6\hatD_b+\cdots,\,\m6\hatD_a+\cdots]=\m6\hatR_{ab}+\cdots\,,\nn
0&=&[\hatD_a,\,\delta_Z(\alpha)]=[\m6\hatD_a+\cdots,\,\m6\hatD_5+\cdots]
=\m6\hatR_{5a}+\cdots\,.
\label{eq:DDcomm}
\end{eqnarray}
The commutativity of the central charge with $\hatD_a$ in the second 
equation implies the $U_Z(1)$-covariance of the latter and 
should hold as a result of 
the identification $\partial_z=\XZ$ on the hypermultiplet. (Confirming this 
directly on the gauge fields is also straightforward.) \ We can obtain 
various relations from these equations (\ref{eq:DDcomm}) by comparing 
the coefficients of each generator $\XX_A$ on both sides. For example, 
the terms proportional to the central charge $\XZ$ in the first 
equation yield
\begin{eqnarray}
\hat F_{ab}(A)=-2\alpha\omega_{[ab]}{}^5,
\end{eqnarray}
where $\hat{\cal R}_{\mu\nu}(Z)$ here is denoted by the more common notation 
$\hat F_{\mu\nu}(A)$.
The terms proportional to $\XP_a$ and ${\mbf Z}$ in the second equation 
give, respectively, 
\begin{equation}
\quad \omega_{a5}{}^b=\omega_{5a}{}^b,\qquad 
\omega_5{}^{a5}=0 .
\end{equation}
Similarly the following relations can be found:
\begin{eqnarray}
\hat R_{5a}{}^i(Q)&=&\myfrac1{16\alpha}\gamma\dt \hat F(A)\psi_a^i+t^i{}_j\psi_a
^j+\myfrac14\gamma\dt v\psi_a^i+\phi^i_a+\gamma_a\phi^i_5\nn
&=&\{\phi^i_a-\eta^i(\psi_a)\}+\gamma_a\phi^i_5\,,\nn
\hat R_{ab}{}^i(Q)&=&\hat {\cal R}_{ab}{}^i(Q)
-2\gamma_{[a}\{\phi_{b]}-\eta^i(\psi_{b]})\}\,,\nn
\hat R_{ab}{}^{cd}(M)&=&\hat{\cal R}_{ab}{}^{cd}(M)
-\myfrac1{2\alpha^2}\hat F_{ab}(A)\hat F^{cd}(A)
-\myfrac1{2\alpha^2}\hat F_{[a}{}^{[c}(A)\hat F_{b]}{}^{d]}(A)\nn
&&{}+8\{f_{[a}{}^{[c}-\xi_K^{[c}(\psi_{[a})\}e_{b]}{}^{d]}\,,\nn
\hat R_{5a}{}^{5b}(M)&=&\myfrac1{4\alpha^2}\hat F_{ac}(A)\hat F^{bc}(A)
+2\{f_a{}^b-\xi_K^a(\psi_b)\}+2f_5{}^5e_a{}^b\,, \nn
\hat R_{5a}{}^{ij}(U)&=&\hatD_at^{ij}\,,\nn
\hat R_{ab}{}^{ij}(U)&=&\hat {\cal R}_{ab}{}^{ij}(U)
 -\myfrac1\alpha\hat F_{ab}(A)t^{ij}\,.
\label{eq:Rrel}
\end{eqnarray}
Putting these into the constraints (\ref{eq:6D.cons}) on 
the 6D curvatures $\m6\hatR_{\mu\nu}{}^{ab}(M)$ 
and $\m6\hatR_{\mu\nu}^{\ \,i}(Q)$,  we obtain
\begin{eqnarray}
\phi_\mu^i-\eta^i(\psi_\mu)&=&
\myfrac1{16}(\gamma^{ab}\gamma_\mu-\myfrac35\gamma_\mu\gamma^{ab})\hat{\cal R}_{ab}{}^i(Q)
-\myfrac1{120}\gamma_\mu\chi^i\,, \nn
\phi_5^i&=&\myfrac1{40}\gamma^{ab}\hat {\cal R}_{ab}{}^i(Q)
-\myfrac1{120}\chi^i\,,\nn
f_a{}^a-\xi_K^a(\psi_a)+f_5{}^5&=&\myfrac1{20}\hat{\cal R}(M)
+\myfrac1{80\alpha^2}\hat F(A)^2-\myfrac1{40}D\,, 
 \end{eqnarray}
\begin{equation}
\hat{\cal R}_{ab}(M)\equiv\hat{\cal R}_{ac}{}^c{}_b(M)\,,
\qquad \hat{\cal R}(M)\equiv\hat{\cal R}_a{}^a(M)\,. 
\end{equation}

\subsection{Supersymmetry transformation law of the Weyl multiplet}

The commutation relation for the 5D 
supersymmetry transformation $\delta_Q(\varepsilon)$ given in Eq.~(\ref{eq:SusyTrf}) 
is now easily found from the superconformal algebra (\ref{eq:6D.algebra})
in 6D with the help of the relations (\ref{eq:Drel}) of covariant 
derivatives:
\begin{eqnarray}
{}[\delta_Q(\varepsilon_1),\ \delta_Q(\varepsilon_2)]
&=&\xi^a\hatD_a+\delta_Z(\alpha\xi^5)+\delta_M(\myfrac1\alpha\xi^5\hat F(A)_{ab}
+\xi_{abcd}v^{cd}+2\xi^{ij}_{ab}t_{ij})\nn 
&&\qquad {}+\delta_U(-3\xi^5t^{ij}-\myfrac1{2\alpha}\xi^{ij}_{ab}\hat F(A)^{ab}
-2\xi^{ij}_{ab}v^{ab}),
\label{eq:QQalgebra}
\end{eqnarray}
where
\begin{equation}
\xi^a\equiv2i\bar\varepsilon_1\gamma^a\varepsilon_2, \quad 
\xi^5\equiv-2i\bar\varepsilon_1\varepsilon_2, \quad 
\xi_{abcd}\equiv2i\bar\varepsilon_1\gamma_{abcd}\varepsilon_2, \quad 
\xi^{ij}_{ab}\equiv2i\bar\varepsilon_1^{(i}\gamma_{ab}\varepsilon_2^{j)}. 
\end{equation}

If we use the relations (\ref{eq:Drel}) and (\ref{eq:Rrel}),
it is now straightforward (although tedious) to obtain, from
(\ref{eq:6DWeylTrf}) and (\ref{eq:6DMTrf}), the following 
supersymmetry transformation law for our Weyl multiplet in 5D. With 
$\delta=\delta_Q(\varepsilon)$,
\begin{eqnarray}
\delta e_\mu{}^a&=& 
-2i\bar\varepsilon\gamma^a\psi_\mu, \nn
\delta\psi_\mu^i&=&\calD_\mu\varepsilon^i+\myfrac12\gamma_{\mu ab}\varepsilon^iv^{ab}
+\myfrac1{2\alpha}\gamma^a\varepsilon^i\hat F(A)_{\mu a}+\gamma_\mu\varepsilon^jt^i{}_j, \nn
\delta\alpha&=&0, \quad\  (\calD_\mu\varepsilon^i\equiv\partial_\mu\varepsilon^i+\half b_\mu\varepsilon^i
-\myfrac14\omega_\mu^{ab}\gamma_{ab}\varepsilon^i-V_\mu^{\ i}{}_j\varepsilon^j) \nn
\delta A_\mu&=& 
2i\alpha\bar\varepsilon\psi_\mu,  \nn 
\delta V_\mu^{ij}&=& 
-4i\bar\varepsilon^{(i}\gamma_\mu\tilde\chi^{j)}
-2i\bar\varepsilon^{(i}\gamma^a{\hat\calR}_{a\mu}{}^{j)}(Q) \nn
&&{}+\myfrac{i}{\alpha}\bar\varepsilon^{(i}\gamma\dt \hat F(A)\psi_\mu^{j)}
+4i\bar\varepsilon^{(i}\gamma\dt v\psi_\mu^{j)}-6i\bar\varepsilon\psi_\mu t^{ij},\nn
\delta t^{ij}&=& 
4i\bar\varepsilon^{(i}\tilde\chi^{j)},\nn
\delta v_{ab}&=&\myfrac{i}4\bar\varepsilon\gamma_{abcd}\hat{\cal R}^{cd}(Q)
-2i\bar\varepsilon\gamma_{ab}\tilde\chi,\nn
\delta\tilde\chi^i&=&-\myfrac12\gamma^a\varepsilon^i\hatD^bv_{ab}
+\myfrac12\slashD t^i{}_j\varepsilon^j-\gamma\dt vt^i{}_j\varepsilon^j-\myfrac1{4\alpha}\gamma\dt
\hat F(A)t^i{}_j\varepsilon^j \nn
&&{}+\myfrac12C\varepsilon^i
-\myfrac1{32\alpha^2}\gamma^{abcd}\varepsilon^i\hat F_{ab}(A)\hat F_{cd}(A),\nn
\delta C&=&-2i\bar\varepsilon\slashD\tilde \chi+22i\bar\varepsilon^i\tilde \chi^jt_{ij}
-3i\bar\varepsilon\gamma\dt v\tilde \chi-i\bar\varepsilon^i\gamma^{ab}\hat{\cal
R}_{ab}{}^j(Q)t_{ij}, \nn
\delta\omega_\mu{}^{ab}&=&
-2i\bar\varepsilon\gamma^{[a}\hat{\cal R}_\mu{}^{b]}(Q)-i\bar\varepsilon\gamma_\mu\hat{\calR}^{ab}(Q)\nn
&&{}+\myfrac{2i}{\alpha}\bar\varepsilon\psi_\mu\hat F^{ab}(A)-2i\bar\varepsilon\gamma^{abcd}\psi_\mu 
v_{cd}-4i\bar\varepsilon^i\gamma^{ab}\psi_\mu^jt_{ij}.
\label{eq:5DWtrf}
\end{eqnarray}
Here $\tilde\chi^i$ and $C$ are the redefined fields from $\chi^i$ and 
$D$ in Eq.~(\ref{eq:deflambda}), and we have also written the 
transformation law of the spin connection, 
although it is a dependent field 
given by (from Eq.~(\ref{eq:6Dspinconn}))
\begin{equation}
\omega_\mu^{\ ab} = \omega_\mu^{0\ ab}+i(2\bar\psi_\mu\gamma^{[a}\psi^{b]} 
    +\bar\psi^a\gamma_\mu\psi^b)-2e_\mu^{\ [a}\alpha^{-1}\partial^{b]}\alpha.
\end{equation}
The $[\delta_Q,\ \delta_Q]$ commutation relation (\ref{eq:QQalgebra}) can also 
be read directly from the structure functions appearing in 
Eq.~(\ref{eq:5DWtrf}): comparing Eqs.~(\ref{eq:Galg}) and 
(\ref{eq:gaugeTrf}), we see that, for instance, the last three terms in 
$\delta\omega_\mu{}^{ab}$ and $\delta V_\mu^{ij}$ just give the $\delta_M$ and $\delta_U$ terms
in Eq.~(\ref{eq:QQalgebra}), respectively.

When deriving these transformation laws (\ref{eq:5DWtrf}), 
we need the following
transformation laws of curvature tensors, which follow from 
the formula (\ref{eq:Rtrf}): 
\begin{eqnarray}
\delta\hat F_{ab}(A)&=&2i\alpha\bar\varepsilon\hat {\cal R}_{ab}(Q), \nn
\delta\hat{\calR}(M)&=&4i\bar\varepsilon\hatD^a\gamma^b\hat{\calR}_{ab}(Q)
+4i\bar\varepsilon\gamma^{abcd}\hat{\calR}_{ab}(Q)v_{cd}
+4i\bar\varepsilon\gamma^{ab}\hat{\calR}_{bc}(Q)v^c{}_a \nn
&&{}+\myfrac{2i}\alpha\bar\varepsilon\gamma^{ab}\hat{\cal R}_{bc}(Q)\hat F^c{}_a(A)
-\myfrac{4i}\alpha\bar\varepsilon\hat{\calR}_{ab}(Q)\hat F^{ab}(A)
+8i\bar\varepsilon^i\gamma\dt\hat{\calR}^j(Q)t_{ij}, \nn
\delta\,\gamma\dt \hat{\calR}^i(Q)&=&2(\hatD_a\gamma\dt v)\gamma^a\varepsilon^i
-10\gamma^a\varepsilon^i\hatD^bv_{ab}-\myfrac1\alpha\gamma^a\varepsilon^i\hatD^b\hat
F_{ab}(A)+8\slashD t^i{}_j\varepsilon^j \nn
&&{}+\gamma^{abcd}\varepsilon^i\bigl(2v_{ab}v_{cd}-\myfrac1\alpha v_{ab}\hat F_{cd}(A)
-\myfrac1{2\alpha^2}\hat F_{ab}(A)\hat F_{cd}(A)\bigr)\nn
&&{}+\myfrac4\alpha\gamma^{ab}\varepsilon^i\hat F_{ac}(A)v_b{}^c
-12\gamma\dt vt^i{}_j\varepsilon^j-\myfrac6\alpha\gamma\dt \hat F(A)t^i{}_j\varepsilon^j
-\gamma\dt \hat{\cal R}^i{}_j(U)\varepsilon^j\nn
&&{}-\bigl(\myfrac12\hat{\cal R}(M)-6v\dt v +\myfrac1{2\alpha^2}\hat F(A)\dt \hat
F(A)+20t^j{}_k t^k{}_j\bigr)\varepsilon^i.
\end{eqnarray}
We also need the Bianchi identities following from Eq.~(\ref{eq:Bianchi}):
\begin{eqnarray}
\hatD_{[a}\hat{\cal R}^{\ i}_{bc]}(Q)&=&
-\myfrac12\gamma_{de[a}\hat{\cal R}^{\ i}_{bc]}(Q)v^{de}
-\myfrac1{2\alpha}\gamma^d\hat{\cal R}^{\ i}_{[ab}(Q)\hat F_{c]d}(A)
-\gamma_{[a}\hat{\cal R}^{\ j}_{bc]}(Q)t^i{}_j, \nn
\hatD_{[a}\hat F_{bc]}(A)&=&0.
\end{eqnarray}

We will not need explicit expressions for our 5D curvatures 
in this paper, but we list them for the reader's convenience:
\begin{eqnarray}
\hat F_{\mu\nu}(A)&=&2\partial_{[\mu}A_{\nu]}-2i\alpha\bar\psi_\mu\psi_\nu\,,\nn
\hat{\cal R}_{\mu\nu}{}^i(Q)&=&2\calD_{[\mu}\psi_{\nu]}^i
+\gamma_{ab[\mu}\psi_{\nu]}^iv^{ab}
-\myfrac1{\alpha}\gamma^a\psi_{[\mu}^i\hat F_{\nu]a}(A)+2\gamma_{[\mu}\psi_{\nu]}^jt^i{}_j\,, \nn
\hat{\cal R}_{\mu\nu}{}^{ab}(M)&=&
2\partial_{[\mu}\omega_{\nu]}{}^{ab}-2\omega_{[\mu}{}^{[ac}\omega_{\nu]c}{}^{b]}
+4i\bar\psi_{[\mu}\gamma^{[a}\hat{\cal R}_{\nu]}{}^{b]}(Q)
+2i\bar\psi_{[\mu}\gamma_{\nu]}\hat{\calR}^{ab}(Q)\nn
&&{}-\myfrac{2i}{\alpha}\bar\psi_{\mu}\psi_{\nu}\hat F^{ab}(A)
+2i\bar\psi_{\mu}\gamma^{abcd}\psi_{\nu}v_{cd}
+4i\bar\psi_{\mu}^i\gamma^{ab}\psi_{\nu}^jt_{ij}\,, \nn
\hat{\cal R}_{\mu\nu}{}^{ij}(U)&=&
2\partial_{[\mu}V_{\nu]}{}^{ij}+2V_{[\mu}{}^{k(i}V_{\nu]}{}^{j)}{}_k
+8i\bar\psi_{[\mu}^{(i}\gamma_{\nu]}\tilde\chi^{j)}
+4i\bar\psi_{[\mu}^{(i}\gamma^a{\hat\calR}_{a\nu]}{}^{j)}(Q)\nn
&&{}-\myfrac{i}{\alpha}\bar\psi_{\mu}^{(i}\gamma\dt \hat F(A)\psi_{\nu}^{j)}
-4i\bar\psi_{\mu}^{(i}\gamma\dt v\psi_{\nu}^{j)}+6i\bar\psi_{\mu}\psi_{\nu}t^{ij}\,. 
\end{eqnarray}
In the following, a quantity ${\cal C}$ appears. This is 
related to $C$ (or $D$) and the scalar curvature $\hat{\cal R}(M)$ as
\begin{eqnarray}
{\cal C}&\equiv&\myfrac16D+4(f_a{}^a-\xi_K^a(\psi_a)+f_5{}^5)\nn
&=&C+\myfrac14\hat{\cal R}(M)
+\myfrac1{8\alpha^2}\hat F(A)^2-v\dt v+2t^i{}_jt^j{}_i\,.   
\end{eqnarray}

\section{Transformation laws of matter multiplets in 5D}

Now it is straightforward to derive the supersymmetry transformation 
rules for the matter multiplets in 5 dimensions from the 
superconformal rules in 6D given by BSVP, which are 
summarized in Appendix B. 
We can simply apply the formulas (\ref{eq:SusyTrf}) and 
(\ref{eq:Drel}) for our supersymmetry transformation $\delta_Q(\varepsilon)$ 
and covariant derivative $\hatD_\mu$.

In the following we omit explicit expressions for the covariant 
derivatives $\hatD_\mu\phi$ on various fields $\phi$ for conciseness.
In our 5D calculus we have
\begin{eqnarray}
\hatD_\mu\phi&=& \calD_\mu\phi-\delta_Q(\psi_\mu)\phi, \nn
\calD_\mu\phi&=& \bigl(\partial_\mu-\delta_M(\omega_\mu^{ab})-\delta_U(V_\mu^{ij})
-\delta_D(\alpha^{-1}\partial_\mu\alpha)-\delta_Z(A_\mu)-\delta_G(W_\mu)\bigr)\phi, 
\hspace{2em}
\end{eqnarray}
where $\calD_\mu$ is covariant only with respect to homogeneous 
transformations $\XM_{ab},\ \XU_{ij}$, $\XD,\ \XZ$ and 
$\XG$ (Yang-Mills gauge transformation).
The transformation rules under such homogeneous 
transformations are obvious, and are the same as those given in 
(\ref{eq:homotrfs}) for the 6D case. The supersymmetry transformation 
rules are explicitly given below for all the fields, so that the covariant 
derivatives $\hatD_\mu\phi$ will be clear.

\subsection{Vector multiplet}

\begin{table}[tb]
\caption{Matter multiplets in 5D.}
\label{table:5DM}
\begin{center}
\begin{tabular}{ccccc}\hline \hline
field      & type      &  restrictions & {\it SU}$(2)$&  Weyl-weight    \\ \hline 
\multicolumn{5}{c}{Vector multiplet} \\ \hline
$W_\mu$      &  boson    & real gauge field   &  \bf{1}    &   0     \\
$M$& boson & real,\quad $M=-W_5$& \bf{1} & 1 \\ 
$\Omega^i$      &  fermion  &{\it SU}$(2)$-Majorana  & \bf{2} &\hspace{-1mm}\myfrac32 \\  
$Y_{ij}$    &  boson    & $Y^{ij}=Y^{ji}=-Y^*_{ij}$   & \bf{3} & 2 \\ \hline
\multicolumn{5}{c}{Linear multiplet} \\ \hline
\Tspan{$L^{ij}$}& boson & $L^{ij}=L^{ji}=-(L_{ij})^*$  &  \bf{3}   & 3 \\ 
$\varphi^i$ &  fermion  & {\it SU}$(2)$-Majorana & \bf{2}&\hspace{-1mm}\myfrac72 \\  
$E_a$ & boson & real,\quad constrained by (\ref{eq:Con.E})& \bf{1} & 4 \\ 
$N$ & boson & real,\quad $N=-E_5$ & \bf{1} & 4 \\ \hline
\multicolumn{5}{c}{Nonlinear multiplet} \\ \hline
\Tspan{$\Phi^i_\alpha$}  &  boson    & $(\Phi^\alpha_i)^*=-\Phi^i_\alpha$  &  \bf{2} & 0 \\ 
$\lambda^i$   &  fermion  & {\it SU}$(2)$-Majorana & \bf{2}    &\hspace{-1mm}\myfrac12 \\  
$V_a$    &  boson    & real & \bf{1} & 1 \\ 
$V_5$   & boson & real & \bf{1} & 1 \\ \hline
\multicolumn{5}{c}{Hypermultiplet} \\ \hline
\Tspan{$\hA_i^\alpha$}     &  boson & 
$\hA^i_\alpha=\epsilon^{ij}\hA_j^\beta\rho_{\beta\alpha}=-(\hA_i^\alpha)^*$ &\bf{2}& \hspace{-0.8mm}\myfrac32 \rule[-1mm]{0pt}{6mm} \\  
$\zeta^\alpha$    &  fermion  & $\bar\zeta^\alpha\equiv(\zeta_\alpha)^\dagger\gamma_0 = \zeta^{\alpha\T}C$ 
& \bf{1}  & 2 \\ 
$\hF_i^\alpha$  &  boson    & 
$\hF^i_\alpha=-(\hF_i^\alpha)^*$  &  \bf{2}   &\hspace{-0.8mm}\myfrac52 \rule[-3mm]{0pt}{3mm}\\ \hline
\end{tabular}
\end{center}
\end{table}

The vector multiplet in 5D derived from the 6D one consists of the fields 
given in Table \ref{table:5DM}, where the real scalar $M$ comes from 
the fifth spatial component of a 6D vector, $M\equiv-W_5=-\alpha W_z$. Note that 
it carries a Weyl weight $w=1$, since $w(\alpha)=1$ and $w(W_{\6\mu})=0$. As in 
6D, all the component fields are Lie-algebra valued, e.g., $M$ is a 
matrix $M^\alpha{}_\beta=M^A(t_A)^\alpha{}_\beta$, where the $t_A$ are (anti-hermitian) 
generators of the gauge group $G$. The \XQ\ transformation rules are found
from (\ref{eq:6DVtrf}) to be
\begin{eqnarray}
\delta W_\mu&=&-2i\bar\varepsilon\gamma_\mu\Omega+2i\bar\varepsilon\psi_\mu M\,, \nn
\delta M&=&2i\bar\varepsilon\Omega\,, \nn
\delta\Omega^i&=&
-\myfrac14\gamma\dt \hatG(W)\varepsilon^i-\myfrac12\gamma^a\varepsilon^i\hatD_aM-Y^{ij}\varepsilon_j\,, \nn
\delta Y^{ij}&=&2i\bar\varepsilon^{(i}\slashD\Omega^{j)}-i\bar\varepsilon^{(i}\gamma\dt v\Omega^{j)}
 +2i\bar\varepsilon^{(i}t^{j)}{}_k\Omega^k+4i\bar\varepsilon\Omega t^{ij}-2ig\bar\varepsilon^{(i}[M, \Omega^{j)}]\,,
\hspace{2em}
\label{eq:Vec.trf}
\end{eqnarray}
where $g$ is a gauge coupling constant and $\hatG_{ab}(W)$ is the following 
combination of the supercovariant field strength $\hat F_{ab}(W)$ of $W_
\mu$ and that of the gravi-photon $A_\mu$:
\begin{equation}
\hatG_{ab}(W)\equiv\hat F_{ab}(W)-\myfrac1\alpha M\hat F_{ab}(A),
\end{equation}
which is actually the field strength $\6{\hat F}_{ab}(W)$ in 6D.
The gauge group $G$ can be regarded as a sub-group of the super
group, and the above transformation law of the gauge field $W_\mu$ 
provides us with the additional structure functions, $f_{PQ}{}^G$ and 
$f_{QQ}{}^G$. For example, the transformation law of this mixed field 
strength can be obtained easily from (\ref{eq:Rtrf}):
\begin{eqnarray}
\delta\hatG_{ab}(W)&=&4i\bar\varepsilon\gamma_{[a}\hatD_{b]}\Omega-2i\bar\varepsilon\gamma_{abcd}\Omega v^{cd}
-4i\bar\varepsilon\gamma_{c[a}\Omega v_{b]}{}^c\nn
&&{}-\myfrac{2i}{\alpha}\bar\varepsilon\gamma_{c[a}\Omega\hat F_{b]}{}^c(A)
-4i\bar\varepsilon^i\gamma_{ab}\Omega^jt_{ij}\,.
\end{eqnarray}
%

\subsection{Linear multiplet}

The linear multiplet consists of the 
components listed in Table \ref{table:5DM} and may generally 
carry a non-Abelian charge of the gauge group $G$. 
A point which should be noted in the reduction in this case is that, for
later convenience in constructing actions, we have lowered the Weyl 
weight of this multiplet in 5D by one from that in 6D by multiplying 
each component field by $\alpha^{-1}$. 

The $\XQ$ transformation rules following from (\ref{eq:6DLtrf}) read
\begin{eqnarray}
\delta L^{ij}&=&2i\bar\varepsilon^{(i}\varphi^{j)}\,, \nn
\delta\varphi^i&=&-\slashD L^{ij}\varepsilon_j-4t^l{}_kL^k{}_l\varepsilon^i-6t^{(i}{}_kL^{j)k}\varepsilon_j
+gML^{ij}\varepsilon_j \nn
&&{}+\myfrac12\gamma^a\varepsilon^iE_a+\myfrac12\varepsilon^iN
+\myfrac1{2\alpha}\gamma\dt \hat F(A)\varepsilon_jL^{ij}+2\gamma\dt v\varepsilon_jL^{ij}\,, \nn
\delta E_a&=&2i\bar\varepsilon\gamma_{ab}\hatD^b\varphi
-\myfrac{i}{\alpha}\bar\varepsilon\gamma_{abc}\varphi\hat F^{bc}(A)
+\myfrac{i}{\alpha}\bar\varepsilon\gamma^b\varphi\hat F_{ab}(A)
-2i\bar\varepsilon\gamma_{abc}\varphi v^{bc}+6i\bar\varepsilon\gamma^b\varphi v_{ab} \nn
&&{}-8i\bar\varepsilon^i\gamma_a\varphi^jt_{ij}
+2i\bar\varepsilon^i\gamma_{abc}\hat {\cal R}^{bcj}(Q)L_{ij} 
+2ig\bar\varepsilon\gamma_aM\varphi-4ig\bar\varepsilon^i\gamma_a\Omega^jL_{ij}\,, \nn
\delta N&=&-2i\bar\varepsilon\slashD\varphi
-3i\bar\varepsilon\gamma\dt v\varphi
-2i\bar\varepsilon^i\gamma\dt\hat{\cal R}^j(Q)L_{ij} 
-10i\bar\varepsilon^i\varphi^jt_{ij}+4ig\bar\varepsilon^i\Omega^jL_{ij}\,.
\hspace{1em}
\label{eq:LinearTrf}
\end{eqnarray}
This multiplet apparently contains nine Bose and eight Fermi fields. 
Thus closure 
of the algebra requires the following $\XQ$-invariant constraint, 
which also follows directly from that in 6D, (\ref{eq:6DLcons}):
\begin{eqnarray}
\hatD^aE_a+i\bar\varphi\,\gamma\dt\hat{\calR}(Q)
+gMN+4ig\bar\Omega\varphi+2gY^{ij}L_{ij}=0.
\label{eq:Con.E}
\end{eqnarray}

When the linear multiplet carries no gauge-group charges 
(i.e., $g=0$ in the above transformation law), the constraint 
(\ref{eq:Con.E}) can be solved with respect to $E^a$. Indeed, the 
constraint can be rewritten in the form $e^{-1}\partial_\lambda(eV^\lambda)=0$ with
\begin{equation}
V^\lambda\equiv e^\lambda_aE^a+2i\bar\psi_\rho\gamma^{\lambda\rho}\varphi
+2i\bar\psi_\mu^i\gamma^{\lambda\mu\nu}\psi_\nu^jL_{ij}.
\end{equation}
Hence there exists an antisymmetric tensor gauge field 
$E^{\mu\nu}$, which is a tensor-density and possesses an additional gauge 
symmetry $\delta E^{\mu\nu}=\partial_\rho(\Lambda^{\mu\nu\rho})$:
\begin{equation}
eV^\lambda=-\partial_\rho E^{\lambda\rho} \quad \rightarrow\quad 
E^a=-e^{-1}e^a_\mu\hatD_\nu E^{\mu\nu},
\label{eq:Erep}
\end{equation}
where $\hatD_\nu E^{\mu\nu}$ is defined by
\begin{equation}
\hatD_\nu E^{\mu\nu}=\partial_\nu E^{\mu\nu}+2ie\bar\psi_\nu\gamma^{\mu\nu}\varphi
+2ie\bar\psi_\nu^i\gamma^{\mu\nu\rho}\psi_\rho^jL_{ij}.
\end{equation}
Because of the covariance of $E^a$, the $\XQ$ transformation law of $E^
{\mu\nu}$ must take the following form, which is also in accordance 
with a direct calculation:
\begin{equation}
\delta E^{\mu\nu}=-2ie\bar\varepsilon\gamma^{\mu\nu}\varphi
-4ie\bar\varepsilon^i\gamma^{\mu\nu\rho}\psi^j_\rho L_{ij}.
\end{equation}
\subsection{Nonlinear multiplet}

The nonlinear multiplet is a multiplet whose component fields are 
transformed nonlinearly. The component fields are also shown in 
Table \ref{table:5DM}. 
The first component, $\Phi^i_\alpha$, carries 
an additional gauge-group {\it SU}$(2)$ index $\alpha\ (=1,2)$, as well as the 
superconformal {\it SU}$(2)$ index $i$.
The index $\alpha$ is also raised and lowered by using 
the invariant tensors $\epsilon^{\alpha\beta}$ and $\epsilon_{\alpha\beta}$, 
$\epsilon^{\alpha\beta}=\epsilon_{\alpha\beta}=i(\sigma_2)_{\alpha\beta}$:
\begin{equation}
\Phi^i_\alpha=\Phi^{i\beta}\epsilon_{\beta\alpha},\qquad \epsilon^{\gamma\alpha}\epsilon_{\gamma\beta}=\delta^\alpha{}_\beta. 
\end{equation}
The field $\Phi^i_\alpha$ takes values in {\it SU}$(2)$ and hence satisfies 
\begin{equation}
\Phi^i_\alpha\Phi^\alpha_i=\delta^i_j,\qquad \Phi^\alpha_i\Phi^i_\beta=\delta^\alpha_\beta.
\end{equation}

The $\XQ$ transformation laws of the nonlinear multiplet, following
from (\ref{eq:6DNLtrf}), are
\begin{eqnarray}
\delta\Phi^i_\alpha&=&2i\bar\varepsilon^{(i}\lambda^{j)}\Phi_{j\alpha}\,, \nn
\delta\lambda^i&=&
-\Phi^i_\alpha\slashD\Phi^\alpha_j\varepsilon^j+\myfrac12\gamma^aV_a\varepsilon^i-\myfrac12V_5\varepsilon^i 
-\myfrac{i}2\gamma^a\varepsilon_j\bar\lambda^i\gamma_a\lambda^j-\myfrac{i}2\varepsilon_j\bar\lambda^i\lambda^j \nn
&&{}-\myfrac{i}8\gamma^{ab}\varepsilon_i\bar\lambda\gamma_{ab}\lambda
+\myfrac1{4\alpha}\gamma\dt \hat F(A)\varepsilon^i+\gamma\dt v\varepsilon^i+3t^i{}_j\varepsilon^j
+gM^\alpha_\beta\Phi^i_\alpha\Phi^\beta_j\varepsilon^j\,, \nn
\delta V_a&=&2i\bar\varepsilon\gamma_{ab}\hatD^b\lambda 
-\myfrac{i}{2\alpha}\bar\varepsilon\gamma_{abc}\lambda{\hat F}^{bc}(A)+2i\bar\varepsilon\gamma^b\lambda v_{ab}\nn
&&{}-i\bar\varepsilon\gamma_a\gamma\dt V\lambda-i\bar\varepsilon\gamma_a\lambda V_5
-2i\bar\varepsilon_i\gamma_a\Phi^i_\alpha\slashD\Phi^\alpha_j\lambda^j
-4ig\bar\varepsilon^i\gamma_a\Omega_j{}^\alpha_\beta\Phi^j_\alpha\Phi^\beta_i\nn
&&{}+4i\bar\varepsilon\gamma_a\tilde\chi-i\bar\varepsilon\gamma_a\gamma^{bc}\hat{\calR}_{bc}(Q)
-2i\bar\varepsilon^i\gamma_a\lambda^jt_{ij}-2ig\bar\varepsilon_i\gamma_a\lambda^jM^\alpha_\beta\Phi^i_\alpha\Phi^\beta_j\,,\nn
\delta V_5&=&2i\bar\varepsilon\slashD\lambda-\myfrac{i}{2\alpha}\bar\varepsilon\gamma\dt \hat F(A)\lambda 
-i\bar\varepsilon\gamma^{ab}\hat {\cal R}_{ab}(Q)+4i\bar\varepsilon\tilde\chi\nn
&&{}-i\bar\varepsilon\gamma\dt V\lambda-i\bar\varepsilon\lambda V_5 +i\bar\varepsilon\gamma\dt v\lambda-2i\bar\varepsilon_i\Phi^i_\alpha 
\slashD\Phi^\alpha_j\lambda^j \nn 
&&{}-2ig\bar\varepsilon_i\lambda^jM^\alpha_\beta\Phi^i_\alpha\Phi^\beta_j
-4ig\bar\varepsilon^i\Omega_j{}^\alpha_\beta\Phi^j_\alpha\Phi^\beta_i\,.
\hspace{3em}
\end{eqnarray}
As in the linear multiplet case, the nonlinear multiplet also needs the
following $\XQ$-invariant constraint for the closure of the algebra:
\begin{eqnarray}
&2{\cal C}+\hatD_aV^a+2i\bar\lambda\slashD\lambda+\hatD^a\Phi^i_\alpha\hatD_a\Phi^\alpha_i
-\myfrac12V^aV_a+\myfrac12V_5^2&\nn
&{}+8i\bar\lambda\tilde\chi-i\bar\lambda\gamma^{ab}\hat{\cal R}_{ab}(Q)
-\myfrac{i}{2\alpha}\bar\lambda\gamma\dt \hat F(A)\lambda+i\bar\lambda\gamma\dt v\lambda 
+2i\bar\lambda^i\Phi_{\alpha i}\slashD \Phi^\alpha_j\lambda^j&\nn
&{}+2gY_{ij}{}^\alpha_\beta\Phi^i_\alpha\Phi^{\beta j}
-8ig\bar\lambda^i\Omega_j{}^\alpha_\beta\Phi^j_\alpha\Phi^\beta_i
+2ig\bar\lambda^i\lambda^jM^\alpha_\beta\Phi_{\alpha i}\Phi^\beta_j&\nn
&{}+t^i{}_jt^j{}_i-2gM_{\alpha\beta}t^{ij}\Phi^\alpha_i\Phi^\beta_j+g^2M^\alpha_\beta M^\beta_\alpha=0.&
\end{eqnarray}

\section{Hypermultiplet}

\subsection{Off-shell hypermultiplet}

The hypermultiplet consisting of a boson $\hA^\alpha_i$ and a fermion $\zeta^\alpha$ 
in 6D is an on-shell multiplet, and thus 
the supersymmetry algebra closes only when it satisfies the 
equations of motion. When we go down to 5D, we can make it an off-shell 
multiplet. The procedure is essentially identical to that known for the 
4D case.\cite{ref:dWLVP,ref:onoff} 
The only difference is that the necessary 
$U_Z(1)$ gauge multiplet is not added by hand but is automatically 
included in our case as a multiplet containing the 
gravi-photon field $A_\mu$.

Explicitly we proceed as follows. In the original 
supersymmetry transformation law (\ref{eq:6DhyperTrf}) of the 
hypermultiplet in 6D, we first make all the Weyl multiplet and 
transformation parameters $z$ independent, 
while keeping only the hypermultiplet members $z$ dependent. Even at 
this stage, clearly, we still have the same form for the supersymmetry 
algebra as the original 6D form (\ref{eq:6D.uncl}), 
in which $\Gamma^\alpha$ is the non-closure function on $\zeta^\alpha$ given in 
(\ref{eq:defGammaC}). However, as shown by Eq.~(\ref{eq:6D.uncl.trs}), 
$\Gamma^\alpha$ still closes under the superconformal transformations, 
together with $C^\alpha_i$, also defined in (\ref{eq:defGammaC}).
We note that in deriving these transformation laws, the only property 
required for $\partial_z$ is that it commutes with all the other 
transformations $\XX_A$. Therefore,
even if {\it the fifth spatial derivative $\partial_z$ is replaced by the 
central charge transformation $\XZ$ everywhere in these},\footnote{%
The other way around, it is also well-known\cite{ref:SuperspZ} to
treat the central charge by introducing an extra coordinate $z$.} 
exactly the same form for the transformation laws will hold, provided that 
$\XZ$ is actually a {\it central} charge. At this stage, the 6D 
supersymmetry transformation becomes
\begin{equation}
\m6\delta_Q(\varepsilon) = \m6\delta^{\rm org}_Q(\varepsilon)\big|_{\partial_z\rightarrow\XZ}\  . \qquad 
(\varepsilon=\varepsilon(x):\ \hbox{$z$-independent})
\label{eq:Zrepldelta}
\end{equation}
However, then, the conditions $\Gamma^\alpha|_{\partial_z\rightarrow\XZ}=0$ and 
$C^\alpha_i|_{\partial_z\rightarrow\XZ}=0$ are no longer the on-shell conditions but become
merely defining equations for the central-charge transformations 
$\XZ \zeta^\alpha$ and $\XZ(\XZ \hA^\alpha_i)$ in terms of the other fields. Since 
the `on-shell condition' $C^\alpha_i=0$ for the boson $\hA^\alpha_i$ is second order
in $\partial_z$, there appears no constraint on the first central charge 
transformation $\XZ \hA^\alpha_i$, so that it defines an auxiliary field 
\begin{equation}
\hF^\alpha_i\equiv\alpha\XZ \hA^\alpha_i,
\end{equation} 
which is necessary for closing the algebra off-shell and balancing the 
numbers of boson and fermion degrees of freedom. The factor of the `dilaton' 
$\alpha$ is included to adjust the Weyl weight of $\hF^\alpha_i$ so as to have 
$w(\hF^\alpha_i)-w(\hA^\alpha_i)=1$, for convenience.

The superconformal transformation $\XX_A$ of $\hF^\alpha_i$ can be found from 
the requirement that $\XZ$ commutes with all $\XX_A$ as $\XX_A\hF^\alpha_i
=\XZ(\XX_A\alpha\hA^\alpha_i)$. This guarantees the central charge property 
$[\XZ,\,\XX_A]=0$ on $\hA^\alpha_i$. This property holds also on $\zeta^\alpha$ and 
$\hF^\alpha_i$ if their central-charge transformations $\XZ \zeta^\alpha$ and 
$\XZ \hF^\alpha_i=\alpha\XZ(\XZ \hA^\alpha_i)$ are defined by $\Gamma^\alpha|_{\partial_z\rightarrow\XZ}=0$ and 
$C^\alpha_i|_{\partial_z\rightarrow\XZ}=0$, as above. This is 
the case because the set of conditions $\Gamma^\alpha=0$ and $C^\alpha_i=0$ is 
invariant under the superconformal transformations.
For later convenience, we note that $\Gamma^\alpha$ contains two $\partial_z$ terms, 
$-i\gamma^z\partial_z\zeta^\alpha-\gamma^{\6a}\gamma^z(\partial_z\hA^\alpha_j)\psi_{\6a}^j$, so that 
$\delta_Z\zeta$ determined by $\Gamma^\alpha|_{\partial_z\rightarrow\XZ}=0$ can be written 
using $\gamma^z\gamma^z=g^{zz}$ in the form
\begin{equation}
\delta_Z\zeta^\alpha= -i(g^{zz})^{-1}\gamma^z
\bigl(\Gamma^\alpha-\gamma^{\6a}\gamma^z(\alpha^{-1}\hF^\alpha_j-\partial_z\hA^\alpha_j)\psi_{\6a}^j\bigr)+
\partial_z\zeta^\alpha.
\end{equation}

Next we fix the gauges of $\XM_{a5}$, $\XS^i$ and $\XK_{\6a}$ as done 
above, and then the supersymmetry transformation $\delta_Q(\varepsilon)$ in 5D is given 
by Eq.~(\ref{eq:SusyTrf}), where $\m6\delta_Q(\varepsilon)$ is the 6D supersymmetry 
transformation with $\partial_z$ replaced by $\XZ$, given in 
(\ref{eq:Zrepldelta}). The
relations between the covariant derivatives in 6D and 5D are given 
exactly by Eq.~(\ref{eq:Drel}). Here also, it is convenient to lower the
Weyl weight of the hypermultiplet in 5D by $1/2$ from that in 6D by 
multiplying each component by $\alpha^{-1/2}$. Doing this, the Weyl weights 
of the hypermultiplet members become those given in Table \ref{table:5DM}. 

The supersymmetry transformation law for the 5D hypermultiplet determined
this way is given by
\begin{eqnarray}
\delta\calA^i_\alpha&=&2i\bar\varepsilon^i\zeta_\alpha\,,\nn
\delta\zeta_\alpha&=&-\slashD \calA_\alpha^i\varepsilon_i+\calF_\alpha^i\varepsilon_i+3t_{ij}\varepsilon^i\calA_\alpha^j
+gM_\alpha{}^\beta\varepsilon^i\calA_{i\beta}
+\myfrac1{4\alpha}\gamma\dt\hat F(A)\varepsilon_i\calA_\alpha^i+\gamma\dt v\varepsilon_i\calA_\alpha^i\,,\nn
\delta\calF^i_\alpha&=&-{2i}\bar\varepsilon^i\bigl(\slashD\zeta_\alpha 
-\myfrac1{4\alpha}\gamma\dt \hat F(A)\zeta_\alpha 
+\myfrac12\gamma\dt v\zeta_\alpha-\myfrac12\calA_\alpha^i\gamma^{ab}\hat{\cal R}_{abi}(Q)\nn
&&\qquad\qquad  
{}+2\calA^i_\alpha\tilde\chi_i-gM_\alpha{}^\beta\zeta_\beta+2g\Omega^i{}_\alpha{}^\beta\calA_{\beta i}\bigr).
\label{eq:5DhyperTrf}
\end{eqnarray}
The central charge transformation law is
\begin{eqnarray}
\alpha\delta_Z(\theta)\calA^i_\alpha&=&\theta\calF^i_\alpha\,,\nn
\alpha\delta_Z(\theta)\zeta_\alpha&=&-\theta\bigl(\slashD\zeta_\alpha 
-\myfrac1{4\alpha}\gamma\dt \hat F(A)\zeta_\alpha 
+\myfrac12\gamma\dt v\zeta_\alpha-\myfrac12\calA_\alpha^i\gamma^{ab}\hat{\cal R}_{abi}(Q)\nn
&&\qquad
{}+2\calA^i_\alpha\tilde\chi_i-gM_\alpha{}^\beta\zeta_\beta+2g\Omega^i{}_\alpha{}^\beta\calA_{\beta i}
\bigr)\,,\nn
\alpha\delta_Z(\theta)\calF^i_\alpha&=&{\theta}\bigl((\hatD^a\hatD_a -{\cal C})\calA_\alpha^i
+4i\bar\zeta_\alpha\tilde\chi^i +4ig\bar\Omega^i{}_\alpha{}^\beta\zeta_\beta 
-2gY^i{}_{j\alpha}{}^\beta\calA^j_\beta-t^i{}_jt^j{}_k\calA^k_\alpha\nn
&&{}+2gt^i{}_jM_\alpha{}^\beta\calA_\beta^j 
-g^2M_\alpha{}^\beta M_\beta{}^\gamma\calA_\gamma^i
 -2t^i{}_j\calF^j_\alpha+2gM_\alpha{}^\beta\calF_\beta^i\bigr)\,.
\hspace{3em}
\label{eq:Ztrf}
\end{eqnarray}
One should note that the covariant derivatives $\hatD_a$ here as well as 
in (\ref{eq:5DhyperTrf}) contain also the covariantization term 
$-\delta_Z(A_a)$ with respect to the central charge, so that these definitions 
of $\delta_Z(\theta)\zeta_\alpha$ and $\delta_Z(\theta)\hF^i_\alpha$ by the second and third equations 
of (\ref{eq:Ztrf}) are recursive. However, they can easily be solved 
algebraically;  for instance, 
the second equation gives
\begin{eqnarray}
\delta_Z(\theta)\zeta_\alpha&=&-\theta{\alpha+\slash{A}\over\alpha^2-A^2}\bigl(\slashD'\zeta_\alpha 
-\myfrac1{4\alpha}\gamma\dt \hat F(A)\zeta_\alpha 
+\myfrac12\gamma\dt v\zeta_\alpha-\myfrac12\calA_\alpha^i\gamma^{ab}\hat{\cal R}_{abi}(Q)\nn
&&\qquad \qquad \qquad 
{}+2\calA^i_\alpha\tilde\chi_i-gM_\alpha{}^\beta\zeta_\beta+2g\Omega^i{}_\alpha{}^\beta\calA_{\beta i}\bigr),
\end{eqnarray}
where $\hatD_a'$ denotes a covariant derivative with the $-\delta_Z(A_a)$ 
term omitted.

\subsection{Hypermultiplet action}

Note that we did not actually have to throw away the $z$ dependence of 
the hypermultiplet in the above 5D supersymmetry
transformation. If $z$ is retained, it merely represents a continuous 
label of an infinite number of copies of the 5D hypermultiplets which 
all transform in the same way.

In 6D, the action $S_0$ for the hypermultiplet was constructed in the form
\begin{equation}
S_0 =  \int d^5x\int dz\,e_6\,\bigl\{
\hA_i^\alpha d_\alpha{}^\beta C_\beta^i
+2(\bar\zeta^\alpha-i\bar\psi^i_\mu\gamma^\mu\hA_i^\alpha)d_\alpha{}^\beta\Gamma_\beta\bigr\},
\label{eq:hypeAction}
\end{equation}
so that the equations of motion give the desired `on-shell' conditions 
$\Gamma^\alpha=C^\alpha_i=0$,
where $e_6$ is the determinant of the sechsbein and 
$d_\alpha{}^\beta$ is a $G$-invariant tensor. This action is fully invariant 
under the original superconformal transformation in 6D.

Here, in the action, let us take all the Weyl multiplet (and 
transformation parameters) to be $z$ independent, while keeping the 
hypermultiplet $z$ dependent and using the original $C_\alpha^i$ and $\Gamma_\alpha$, 
in which $\partial_z$ are {\it not} replaced by $\XZ$. Then, the
action represents an action for the infinite copies of the 5D 
hypermultiplets labeled by $z$, but is, of course, not invariant under 
the above 6D supersymmetry transformation $\m6\delta_Q(\varepsilon)$ with $\partial_z$ 
replaced by $\XZ$, since, for instance, it contains no auxiliary field 
$\hF_i^\alpha$. However, we can make it invariant with a small modification 
as follows.

From our knowledge of global supersymmetric theory, we expect that 
the \linebreak quadratic term $\partial_z\hA^\alpha_id_\alpha{}^\beta\partial_z\hA_\beta^i$ does not 
appear in the action and that the auxiliary field $\hF_i^\alpha$ 
appears as a replacement of $\alpha\partial_z\hA_i^\alpha$. Thus we are led to trying 
the following action (before doing the overall rescaling of the 
hypermultiplet by the factor $\alpha^{-1/2}$): 
\begin{equation}
S =  S_0 + \int d^5x\int dz\,\bigl\{
-e_6g^{zz}(\alpha^{-1}\hF^\alpha_i-\partial_z\hA^\alpha_i)
d_\alpha{}^\beta(\alpha^{-1}\hF^i_\beta-\partial_z\hA^i_\beta)
\bigr\}.
\label{eq:hyperPreAction}
\end{equation}
Indeed, then the added quadratic term in $\alpha^{-1}\hF^\alpha_i-\partial_z\hA^\alpha_i$ 
exactly cancels the quadratic term 
$+e_6g^{zz}\partial_z\hA^\alpha_id_\alpha{}^\beta\partial_z\hA_\beta^i$, which is contained in the 
first term $e_6\hA_i^\alpha d_\alpha{}^\beta C_\beta^i$ in $S_0$ after a partial 
integration. To show that this action $S$ is indeed invariant under the 
the above supersymmetry transformation $\m6\delta_Q(\varepsilon)$ (\ref{eq:Zrepldelta}),
we note that $\m6\delta_Q(\varepsilon)$ transformations of $\tilde
\hF^\alpha_i\equiv\alpha^{-1}\hF^\alpha_i-\partial_z\hA^\alpha_i=(\delta_Z-\partial_z)\hA^\alpha_i$ and 
$e_6g^{zz}$ are given by (in 6D spinor notation)
\begin{eqnarray}
\m6\delta_Q(\varepsilon)\tilde \hF^\alpha_i
&=& 2\bar\varepsilon_i(\delta_Z-\partial_z)\zeta^\alpha=-2i(g^{zz})^{-1}\bar\varepsilon_i\gamma^z
     (\Gamma^\alpha-\gamma^{\6a}\gamma^z\psi_{\6a}^j\tilde\hF^\alpha_j)\,,\nn
\m6\delta_Q(\varepsilon)(e_6g^{zz}) &=& 
-2ie_6g^{zz}\bar\varepsilon\gamma^{\6a}\psi_{\6a}+4ie_6\bar\varepsilon\gamma^z\psi^z
=2ie_6\bar\varepsilon\gamma^z\gamma^{\6a}\gamma^z\psi_{\6a}\,,
\end{eqnarray}
and that the difference between the above $\m6\delta_Q(\varepsilon)$ transformation 
and the original 6D supersymmetry transformation $\m6\delta_Q^{\rm org}(\varepsilon)$
exists only in the $\zeta^\alpha$ field:
\begin{eqnarray}
\m6\delta_Q(\varepsilon)&=& \m6\delta_Q^{\rm org}(\varepsilon) + \delta_Q'(\varepsilon)\,, \nn
\delta'(\varepsilon)\zeta^\alpha&=& i\gamma^z\varepsilon^i\tilde \hF^\alpha_i\,, \quad 
\delta'(\varepsilon)\hA^\alpha_i = 0\,, \quad 
\delta'(\varepsilon)\hbox{(Weyl multiplet)} = 0\,. 
\hspace{1em}
\end{eqnarray}
If we use these equations together with the original action invariance 
$\m6\delta_Q^{\rm org}(\varepsilon)S_0=0$ 
and $\delta S_0/\delta\bar\zeta^\alpha= 4e_6d_\alpha{}^\beta\,\Gamma_\beta$, we can immediately 
confirm the invariance of the action (\ref{eq:hyperPreAction}) 
under $\m6\delta_Q(\varepsilon)$. Use has also been made of the relation
$\tilde\hF^\alpha_jd_\alpha{}^\beta\tilde\hF_\beta^i=
\half\delta^i_j\tilde\hF^\alpha_kd_\alpha{}^\beta\tilde\hF_\beta^k$.

Since the action (\ref{eq:hyperPreAction}) is invariant under 
$\m6\delta_Q(\varepsilon)$, as well as all the other 6D superconformal transformations 
with $z$-independent parameters, 
it gives, after fixing the gauges of $\XM_{a5}$, $\XS^i$ and 
$\XK_{\6a}$, an action that is invariant under the 5D supersymmetry 
transformation $\delta_Q(\varepsilon)$ given by Eq.~(\ref{eq:SusyTrf}).
Invariance under the central charge transformation $\delta_Z$ also 
follows from the $[\delta_Q,\,\delta_Q]$ algebra (\ref{eq:QQalgebra}) and 
invariance under all transformations other than $\delta_Z$.

The action (\ref{eq:hyperPreAction}) contains an infinite number of 
copies of the hypermultiplets. However, we can compactify the 
$z$ direction into a torus of radius $R$ and Fourier expand the 
hypermultiplet fields $\phi^\alpha=(\calA^\alpha_i,\,\zeta^\alpha,\,\calF_i^\alpha)$ into 
cosine and sine modes as 
$\phi^\alpha(x,z)=\sum_n(\phi^{(n)\alpha}_{\c}(x)\cos(nz/R)
+\phi_\s^{(n)\alpha}(x)\sin(nz/R))$. Then, clearly, each set of components, 
$\phi^{(n)\alpha}_\c$ and $\phi^{(n)\alpha}_\s$, with the label $n$ separately 
gives a 5D hypermultiplet closed under the 5D supersymmetry 
transformation, and moreover, the multiplets with different labels $n$ 
are also separated in the action (\ref{eq:hyperPreAction}), which is 
(homogeneously) quadratic in $\phi$, as a result of the conservation of 
momentum $p_z = n/R$. The cosine and sine modes with the same label $n$ 
mix with each other in the terms containing $\partial_z\phi$ in the 
action. Therefore we can retain only the modes with an arbitrary single 
label $n$ to be consistent with invariance. The terms containing no $\partial 
_z$ give the same forms of kinetic terms for the cosine, $\phi^{(n)}_\c$, 
and sine, $\phi^{(n)}_\s$, modes (actually, also independent of $n$), and the
terms containing $\partial_z\phi$ gives mass terms between $\phi^{(n)}_\c$ and $\phi 
^{(n)}_\s$. Since the mass square is $m^2=(n/R)^2$ and $R$ is a free 
parameter, the kinetic terms and the mass terms give separately 
invariant actions. Thus, taking also account of the overall rescaling of
the hypermultiplet by the factor $\alpha^{-1/2}$, we first find the action 
formula for the kinetic terms of the hypermultiplet, which takes the 
same form as the $z$-independent $n=0$ mode,
\begin{eqnarray}
S_{\rm kin} &=&  \int d^5x\,e\,\bigl\{
\hA_i^\alpha d_\alpha{}^\beta C_\beta^{\prime\, i}
+2(\bar\zeta^\alpha-i\bar\psi^i_\mu\gamma^\mu\hA_i^\alpha)d_\alpha{}^\beta\Gamma'_\beta 
\nn && \qquad \qquad \quad 
-g^{zz}(\alpha^{-1}\hF^\alpha_i-\partial_z\hA^\alpha_i)d_\alpha{}^\beta 
(\alpha^{-1}\hF^i_\beta-\partial_z\hA^i_\beta)
\bigr\},\hspace{2em}
\label{eq:kinAction}
\end{eqnarray}
where the prime on $C_\beta^{\prime\, i}$ and $\Gamma'_\beta$ is a reminder that 
the $\partial_z$ terms in them are omitted, although they vanish automatically, 
because the fields are now $z$-independent 5D fields. The mass terms, as 
they stand, are the transition mass terms between the two 
hypermultiplets $\phi^{\alpha}_c$ and $\phi^{\alpha}_s$ (suppressing the label 
$n$). However, if there is a {\it symmetric} $G$-invariant tensor $\eta_ 
{\alpha\beta}=\eta_{\beta\alpha}$, then we can reduce them to a single hypermultiplet by
imposing the constraint
\begin{equation}
\phi_s^\alpha= (d^{-1})^{\alpha\beta}\eta_{\beta\gamma}\phi_c^\gamma\,.
\label{eq:constraint}
\end{equation}
This constraint is consistent with $G$-invariance, and hence with 
supersymmetry. The terms containing $\partial_z$ in the action 
(\ref{eq:hyperPreAction}) have the form $\phi\partial_z\phi$, which is rewritten, 
after substituting $\phi^\alpha(x,z)=\phi^\alpha_c(x)\cos(nz/R)+\phi^\alpha_s(x)\sin(nz/R)$, 
performing the $z$ integration, and imposing the constraint 
(\ref{eq:constraint}), as
\begin{equation}
\phi\partial_z\phi\ \rightarrow\ (n/R)\ \bigl(
\phi_c^\alpha d_{\alpha\beta}\phi_s^{\beta}-\phi_s^{\alpha}d_{\alpha\beta}\phi_c^{\beta}\bigr)=
2(n/R)\ \bigl(\phi_c^{\alpha}\eta_{\alpha\beta}\phi_c^{\beta}\bigr)\,.
\end{equation}
Using this rule and collecting the $\phi\partial_z\phi$ terms in the action 
(\ref{eq:hyperPreAction}), we find the action formula for the 
hypermultiplet mass term to be
\begin{eqnarray}
S_{\rm mass} &=& \int d^5x\,e\, m\eta^{\alpha\beta}\bigl\{
-A^{\mu}(\6{\calD}_\mu\calA_{\alpha i})\calA^i_\beta 
-i\bar\zeta_\alpha\gamma^z\zeta_\beta+2\calA_{\alpha i}
\bar{\6\psi}^i_{\6\mu}\gamma^z\gamma^{\m6\mu}\zeta_\beta\nn
&&\hspace{7em}{} 
-i\bar{\6\psi}^{(i}_{\mu}\gamma^{\mu z\nu}\6\psi^{j)}_{\nu}\calA_{\alpha i}\calA_{\beta j}
+\alpha^{-1}g^{zz}\calF_{\alpha i}\calA^i_\beta\bigr\},
\label{eq:massAction}
\end{eqnarray}
where $\6{\calD}_\mu$ is the covariant derivative with respect to the 
homogeneous transformations $\XM_{\6a\6b},\,\XD,\,\XU^{ij}$ and $\XG$. 
The action formulas (\ref{eq:kinAction}) and (\ref{eq:massAction}) are 
written using the 6D notation (for, in particular, the spinors and 
covariant derivatives), and are generally valid independently of the 
choice of the $\XM_{a5}$, $\XS^i$ and $\XK_{\6a}$ gauge-fixing conditions.
More explicit expressions for them, valid in the present gauge-fixing 
and completely written in 5D notation, will be given in the forthcoming
paper.\cite{ref:KO2}

\section{Embedding and invariant action formulas}

\subsection{Embedding formulas}

We now give some embedding formulas that give a (known type of) 
multiplet using a (set of) multiplet(s). First, however, we discuss the 
important point that there is a vector submultiplet in our 5D Weyl 
multiplet. 

When going down to 5D from 6D, there appeared a 
new set of fields $A_\mu,\ \psi_5$ and $\alpha$ from the fifth spatial components 
of the vielbein and Rarita-Schwinger fields in 6D. It is natural to wonder 
if they might give a submultiplet in the 5D Weyl multiplet. Indeed this 
is the case, and we can easily check that 
the gravi-photon $A_\mu$ and the dilaton $\alpha=e_5^z$ have the same 
transformation law with the usual matter $U(1)$ vector multiplet with 
the identification
\begin{equation}
(W_\mu,\,M,\, \Omega^i,\,Y^{ij})=(A_\mu,\,\alpha,\,0,\,0).
\label{eq:AtoV}
\end{equation}
We suspect that $\psi_5$ would have appeared as the $\Omega$ component of this
multiplet if $\psi_5$ were not set to zero as the $\XS$ gauge-fixing. 
We refer to this vector multiplet as a `central charge vector multiplet'.

Now we give an embedding formula of the vector multiplets ${\mbf V}^\gA$ 
into a linear multiplet. The index $\gA$ labels the generators 
$\{ t_\gA \}$ of the gauge group $G$, which is generally non-simple. 
This formula exists for arbitrary polynomials 
$f(M)$ of the first components $M^A$ of ${\mbf V}^A$, as long as 
the degree of $f$ is two or less:
\begin{equation}
f(M)=f_0+f_{0\gA }M^\gA +\myfrac12f_{0\gA \gB }M^\gA M^\gB.
\end{equation}
For such a polynomial $f$, we can identify the following linear multiplet:
\begin{eqnarray}
L^{ij}&=&-2ft^{ij}+f_\gA Y^{\gA\,ij}
-if_{\gA \gB }\bar\Omega^{\gA i}\Omega^{\gB j}\,,\nn
\varphi^i&=&-4f\tilde \chi^i
+f_\gA \left((\slashD-\myfrac12\gamma\dt v)\Omega^{\gA i}
+t^i{}_j\Omega^{\gA j}-g[M, \Omega^i]^{\gA}\right)\nn
&&{}-f_{\gA \gB }\left( (\myfrac14\gamma\dt \hatG(W)^\gA 
-\myfrac12\slashD M^\gA)\Omega^{\gB i} + Y^{\gA\,i}{}_j\Omega^{\gB j} \right),\nn
E_a&=&\hatD^b(4fv_{ab}+f_\gA \hatG_{ab}(W)^\gA 
+if_{\gA \gB }\bar\Omega^\gA \gamma_{ab}\Omega^\gB )\nn
&&{}-if_\gA \bar\Omega^\gA \gamma_{abc}\hat{\cal R}^{bc}(Q)
+\myfrac1{4\alpha^2}f_0\epsilon_{abcde}\hat F^{bc}(A)\hat F^{de}(A)\nn
&&{}+\myfrac1{4\alpha}\epsilon_{abcde}f_{0\gA} \hat F^{bc}(W)^\gA \hat F^{de}(A) 
+\myfrac18\epsilon_{abcde}f_{0 \gA \gB }
\hat F^{bc}(W)^\gA \hat F^{de}(W)^\gB \nn
&&{}-2igf_\gA [\bar\Omega,\gamma_a\Omega]^\gA 
-2igf_{\gA \gB }\bar\Omega^\gA \gamma_a[M,\Omega]^\gB +gf_\gA [M,\hatD_aM]^\gA\,,\nn
N&=&-4f(C+4t^i{}_jt^j{}_i) -\hatD^a\hatD_a f \nn
&&{}+f_\gA \bigl( 
 -2\hatG_{ab}(W)^\gA v^{ab}
+\myfrac1{2\alpha}\hatG_{ab}(W)^\gA \hat F^{ab}(A)+4t^i{}_jY^{\gA\,j}{}_i \nn
&&\qquad \qquad {}+i\bar\Omega^\gA \gamma^{ab}\hat{\cal R}_{ab}(Q)
-16i\bar\Omega^\gA \tilde\chi+2ig[\bar\Omega,\Omega]^\gA 
\bigr) 
\nn
&&{}+f_{\gA\gB }\bigl(%
-\myfrac14\hatG(W)^\gA\dt \hatG(W)^\gB +\myfrac12\hatD^aM^\gA \hatD_aM^\gB 
-Y^{\gA\,i}{}_j Y^{\gB\,j}{}_i+2i\bar\Omega^\gA\slashD\Omega^\gB \nn
&&\qquad \qquad {}+\myfrac{i}{2\alpha}\bar\Omega^\gA \gamma\dt \hat F(A)\Omega^\gB
 -i\bar\Omega^\gA \gamma\dt v\Omega^\gB +2i\bar\Omega^{\gA i}\Omega^{\gB j}t_{ij}
\bigr), 
\label{eq:VtoL}
\end{eqnarray}
where the commutator $[X,Y]^\gA$ represents $[X,Y]^\gA t_\gA\equiv 
X^\gB Y^\gC [t_\gB,\,t_\gC]$, and
\begin{equation}
f\equiv f(M), \quad f_\gA \equiv{\partial f\over\partial M^\gA }=f_{0\gA } + f_{0\gA \gB }M^\gB,
\quad f_{\gA \gB }\equiv{\partial^2f\over\partial M^\gA \partial M^\gB }=f_{0\gA \gB }.
\end{equation}
When the transformation of the lowest component $L^{ij}$ of 
a linear multiplet takes the form $2i\bar\varepsilon^{(i}\varphi^{j)}$, then 
the supersymmetry algebra (\ref{eq:QQalgebra}) demands that all the 
other higher components must uniquely transform in the form given in 
Eq.~(\ref{eq:LinearTrf}) and that the constraint (\ref{eq:Con.E}) should 
hold. Therefore, in order to identify all the above components of 
the linear multiplet, we have only to examine the transformation law up 
to the second component, $\varphi^i$, since the supersymmetry algebra is
guaranteed to hold for any function of the vector multiplet fields. The 
transformation laws of the remaining components $E^a$ and $N$, as well 
as the constraint, are automatic and need not be checked. 

For a more general function $f(M)$, we cannot satisfy the first 
component transformation form $\delta L^{ij}=2i\bar\varepsilon^{(i}\varphi^{j)}$. 
Therefore, this embedding is impossible for functions 
other than the polynomial $f(M)$ of degree two.

In the above derivation we have assumed that the coefficients 
$f_0,\,f_{0\gA }$ and $f_{0\gA \gB }$ are constants. But actually $f(M)$ 
should carry Weyl weight $w=2$. 
Thus $f_0,\,f_{0\gA }$ and $f_{0\gA \gB }$ in fact each takes the form 
const\,$\times(\alpha^2,\,\alpha,\,1)$ (and hence 
$f(M)$ is actually a homogeneous quadratic polynomial in $\alpha$ and $M$). 
This is consistent since $\alpha$ is covariantly constant and 
$\XQ$ invariant.

If $f(M)$ is $G$-invariant, the above linear multiplet does not 
carry any charge. Then, it must be possible to rewrite the embedded 
$E^a$ component into the form (\ref{eq:Erep}) in terms of an 
antisymmetric tensor gauge field $E^{\mu\nu}$. (Note that the last three 
terms in $E_a$ of Eq.~(\ref{eq:VtoL}) cancel and vanish 
for $G$-invariant $f(M)$.) \ In this case, we find 
\begin{eqnarray}
E^{\mu\nu}&=&-e(4fv^{\mu\nu}+f_\gA \hatG^{\mu\nu}(W)^\gA +if_{\gA \gB }\bar\Omega^\gA \gamma^{\mu\nu}\Omega^\gB 
 +if\bar\psi_\rho\gamma^{\mu\nu\rho\sigma}\psi_\sigma-2if_\gA \bar\psi_\lambda\gamma^{\mu\nu\lambda}\Omega^\gA )\nn
&&{}-\myfrac1{2\alpha^2}f_0\epsilon^{\mu\nu\lambda\rho\sigma}A _\lambda F_{\rho\sigma}(A)
-\myfrac1{2\alpha}f_{0\gA }\epsilon^{\mu\nu\lambda\rho\sigma}A_\lambda F_{\rho\sigma}(W)^\gA \nn
&&{}-\myfrac12f_{0\gA \gB }\epsilon^{\mu\nu\lambda\rho\sigma}(W^\gA _\lambda\partial_\rho W^\gB _\sigma 
   -\myfrac13gW^\gA _\lambda[W_\rho,\,W_\sigma]^\gB )\,.
\label{eq:Eemb}
\end{eqnarray}

We next consider construction of a linear multiplet from the product of two
hypermultiplets $(\calA_\alpha^i,\,\zeta^\alpha,\,\calF_\alpha^i)$ and $({\calA'}_\alpha 
^i,\,{\zeta'}^\alpha,\,{\calF'}_\alpha^i)$. This possibility is suggested by
$N=2$, $d=4$ superconformal tensor calculus.\cite{ref:dWLVP} 
Actually, we find almost the same 
form of formula to hold:
\begin{eqnarray}
L^{ij}&=&\eta^{\alpha\beta}\calA_\alpha^{(i}{\calA'}_\beta^{j)}\,,\nn
\varphi^i&=&\eta^{\alpha\beta}\left(\zeta_\alpha{\calA'}_\beta^i+\calA_\alpha^i\zeta'_\beta\right),\nn
E_a&=&\eta^{\alpha\beta}\left\{\calA_\alpha^i\hatD_a{\calA'}_{\beta i}
   -(\hatD_a\calA_\alpha^i){\calA'}_{\beta i}-2i\bar\zeta_\alpha\gamma_a\zeta'_\beta\right\},\nn
N&=&\eta^{\alpha\beta}\left\{-\calA_\alpha^{i}(gM*{\calA'}_i)_\beta 
  +(gM*\calA^i)_\alpha\calA'_{\beta i}
  -2t_{ij}\calA_\alpha^{i}{\calA'}_\beta^{j}-2i\bar\zeta_\alpha\zeta'_\beta\right\},\nn
&& \left((gM*\hA_i)^\alpha\equiv(\delta_G(M)+\delta_Z(\alpha))\hA_i^\alpha 
= gM^\alpha_{\ \beta}\hA^\beta_i+\hF_i^\alpha\right)
\label{eq:HtoL}
\end{eqnarray}
where $\eta^{\alpha\beta}$ is an arbitrary $G$-covariant tensor. 
For instance, if $\eta^{\alpha\beta}$ is proportional to the 
generator matrices $(t_A)^{\alpha\beta}=\rho^{\beta\gamma}(t_A)^\alpha{}_\gamma$, 
then this linear multiplet belongs to the adjoint representation of 
$G$. Note that even in the case that $\eta^{\alpha\beta}$ is a $G$-invariant 
tensor, this linear multiplet still carries a $U(1)$ charge, i.e., 
it is not invariant under the $\XZ$ transformation; e.g., 
$\delta_Z(\alpha) L^{ij}=\eta^{\alpha\beta}(\hF^{(i}_\alpha\hA^{\prime j)}_\beta 
+\hA^{(i}_\alpha\hF^{\prime j)}_\beta)$. For this linear multiplet, 
therefore, the `group action terms' like $gML^{ij}$ appearing in the 
supersymmetry transformation law (\ref{eq:LinearTrf}) should be 
understood to contain not only the usual gauge group $G$ action but also
the central charge $\XZ$ action; that is, noting that the vector 
multiplet associated with the central charge is the multiplet 
(\ref{eq:AtoV}), the $gM$ action should be understood to be $gM*\equiv 
\delta_G(M)+\delta_Z(\alpha)$, the same action as on $\hA^\alpha_i$ in the above.

\subsection{Invariant action formulas}

An invariant action formula exists for the product of a vector multiplet
and a linear multiplet in 6D, as given in Eq.~(\ref{eq:VLaction}). This 
leads directly to the following invariant action formula in 5D:
\begin{eqnarray}
e^{-1}{\cal L}_{\rm VL}
&=&Y^{ij}L_{ij}+2i\bar\Omega\varphi+2i\bar\psi^i_a\gamma^a\Omega^jL_{ij} 
+\myfrac12M(N-2i\bar\psi_b\gamma^b\varphi
  -2i\bar\psi^i_a\gamma^{ab}\psi^j_b L_{ij}) \nn
&&{}
-\myfrac12W_a(E^a-2i\bar\psi_b\gamma^{ba}\varphi
  +2i\bar\psi^i_b\gamma^{abc}\psi^j_cL_{ij}) \,.
\label{eq:Act.LV}
\end{eqnarray}
As in 6D, this 5D action is invariant if the vector multiplet is abelian
and the linear multiplet carries no gauge group charges or is charged 
only under the abelian group of this vector multiplet. 
When the linear multiplet carries no charges at all, the constrained 
vector field $E^a$ can be replaced by the unconstrained anti-symmetric 
tensor gauge field $E^{\mu\nu}$, as shown in (\ref{eq:Erep}). 
Using this, the second line of the above action (\ref{eq:Act.LV}) can be 
rewritten, up to a total derivative, as
\begin{equation}
-\half e\, W_a(E^a-2i\bar\psi_b\gamma^{ba}\varphi
  +2i\bar\psi^i_b\gamma^{abc}\psi^j_cL_{ij})
\ \ \rightarrow\ \ +\myfrac14F_{\mu\nu}(W)E^{\mu\nu}.
\label{eq:Act.E}
\end{equation}

In five dimensions, this formula also leads to a simpler 
invariant action formula. That is, we have a special vector multiplet 
(\ref{eq:AtoV}) in 5D which we call the central charge vector multiplet. 
We can apply (\ref{eq:Act.LV}) to this vector multiplet. 
We then obtain 
\begin{equation}
(\alpha e)^{-1}{\cal L}_{\rm L}
=N-2i\bar\psi_b\gamma^b\varphi
  -2i\bar\psi^i_a\gamma^{ab}\psi^j_b L_{ij}
  -\myfrac1\alpha A_a(E^a-2i\bar\psi_b\gamma^{ba}\varphi
  +2i\bar\psi^i_b\gamma^{abc}\psi^j_cL_{ij}).
\label{eq:ActL}
\end{equation}
We may call this a linear multiplet action formula, and 
essentially the same formula was found by Zucker.\cite{ref:Zucker} \ 
Again, 
when the linear multiplet carries no charge at all, the above rewriting 
(\ref{eq:Act.E}) is of course possible, and the formula becomes extremely 
simple:
\begin{equation}
{\cal L}_{\rm L}
=e\alpha(N-2i\bar\psi_b\gamma^b\varphi
  -2i\bar\psi^i_a\gamma^{ab}\psi^j_b L_{ij}) +\half F_{\mu\nu}(A)E^{\mu\nu}.
\label{eq:ActL'}
\end{equation}

These action formulas can be used to write the action for a general 
matter-Yang-Mills system coupled to supergravity. If we use 
the above embedding formula (\ref{eq:VtoL}) of vector multiplets into a 
linear multiplet and apply the last linear multiplet action formula 
(\ref{eq:ActL'}), then we obtain a general Yang-Mills-supergravity 
action. If we use the `hypermultiplet$\times$hypermultiplet $\rightarrow$ 
linear multiplet' 
formula (\ref{eq:HtoL}) and apply the linear multiplet action formula 
(\ref{eq:ActL}), then we obtain the action for a general hypermultiplet 
matter system.\footnote{Actually, these two actions for vector 
multiplets and hypermultiplets do not separately give a consistent 
supergravity action, but they do when combined. This is discussed in a 
forthcoming paper.\cite{ref:KO2}} 
The kinetic term for the hypermultiplet $({\calA}_
\alpha^i,\,{\zeta}^\alpha,\,{\calF}_\alpha^i)$ can be obtained if, when using the 
formula (\ref{eq:HtoL}), we take the central-charge transformed 
hypermultiplet $\XZ({\calA}_\alpha^i,\,{\zeta}^\alpha,\,{\calF}_\alpha^i)$ as 
$({\calA'}_\alpha^i,\,{\zeta'}^\alpha,\,{\calF'}_\alpha^i)$. The mass term can be 
obtained by choosing $({\calA'}_\alpha^i,\,{\zeta'}^\alpha,\,{\calF'}_\alpha^i)= 
({\calA}_\alpha^i,\,{\zeta}^\alpha,\,{\calF}_\alpha^i)$ and a symmetric tensor $\eta^ {\alpha 
\beta}$. \cite{ref:dWLVP} It is interesting to see that these formulas give
the same hypermultiplet actions as those which we have independently 
derived from the 6D action in \S4.


\section{Summary and discussion}

In this paper we have derived supergravity tensor calculus in five 
dimensions using dimensional reduction from the known superconformal tensor
calculus in six dimensions. Our 5D supergravity tensor calculus results 
from that in 6D by fixing the gauges of the $\XM_{a5}$, $\XS^i$ and 
$\XK_{\6a}$ symmetries, and so it retains 
the supersymmetry $\XQ^i$, the local Lorentz symmetry $\XM_{ab}$, 
the dilatation symmetry $\XD$ and the gauge symmetries of 
{\it SU}$(2)$ $\XU_{ij}$, 
central charge $\XZ$ and a group $\XG$ transformations. 
We have derived supersymmetry transformation laws for the 
vector multiplet, linear multiplet, nonlinear multiplet and 
hypermultiplet. In particular, we have made the hypermultiplet off-shell, 
while it existed only as an on-shell multiplet in 6D. 

Moreover, we have obtained invariant action formulas and multiplet-embedding
formulas, which will become useful when constructing a general 
matter-gauge field system coupled to supergravity. Some of these 
formulas are derived directly from the 6D formulas through dimensional 
reduction, but others are particular to the present five dimensions. 

The presence of the dilatation $\XD$ in our calculus is extremely useful,
because it makes obtaining the canonical form of the Einstein 
and Rarita-Schwinger terms trivial. Usually, when we write a general 
system of matter and gauge fields coupled to supergravity, the 
Einstein and Rarita-Schwinger terms in the resultant action have a 
non-trivial function of fields as their common coefficient. But if we 
have dilatation symmetry, the coefficient function can be set equal to 1 
simply as a gauge fixing condition of the $\XD$ symmetry.\cite{ref:KU2} 

In this respect, it might have been better not to gauge-fix the $\XS^i$ 
supersymmetry either, since the $\XS^i$ symmetry could be used to 
eliminate the mixing of the Rarita-Schwinger and matter 
fermions.\cite{ref:KU2} However, we have chosen to fix $\XS^i$ by $\psi_z=0$ 
in this paper to avoid complications in the dimensional reduction. The price 
we pay for doing this, however, is that
we need to make some field redefinitions to eliminate the fermion 
mixing. It is also interesting to note that it is actually possible to 
avoid fixing the gauge altogether (even the $\XM_{a5}$ gauge) in the 
course of the dimensional reduction (i.e., making all the fields 
$z$ independent). Then, the resultant 5D theory would have full 6D 
superconformal symmetry aside from the fact that the 6D GC transformation 
reduces to the 5D GC transformation plus the central charge 
transformation. This is interesting because there seems to be no (global) 
superconformal group in 5D. 

By using the formulas given in this paper, it is now easy to write down 
such a general matter-gauge field system coupled to supergravity. To 
obtain such supergravity actions in the superconformal framework, we 
generally need compensating multiplets\cite{ref:KU,ref:VP} 
in addition to the 40+40 
supergravity Weyl multiplet. Our five dimensional theory does not have 
full superconformal symmetries, but its Weyl multiplet is the same size,
and it shares common properties with the superconformal theory. A 
different choice of the compensating multiplets leads to a different 
off-shell formulation of supergravities. In the $N=2$, $d=4$ 
superconformal case, there are three possibilities for the compensator.
The first is to use a nonlinear multiplet,\cite{ref:dWvHVP} the second a
hypermultiplet,\cite{ref:dWvHVP} and the third a linear 
multiplet.\cite{ref:dWPVP} \ In the $d=6$ case, only the third possibility 
is possible, as described in BSVP. On the other hand, here in our 5D 
case, all three possibilities are possible. The first possibility 
was investigated by Zucker in his tensor calculus framework. However, our
experience with $N=1$\cite{ref:FGKVP} and $N=2$\cite{ref:dWLVP} in $d=4$ 
leads us to believe that the second possibility is the most useful 
choice for constructing the most general matter (hypermultiplet) system 
as well as for studying the symmetries. We shall carry out these tasks 
explicitly in a forthcoming paper.\cite{ref:KO2} 

\section*{Acknowledgements}
The authors would like to thank Antoine Van Proeyen for providing them
with precious information on six-dimensional superconformal tensor
calculus. They also appreciate the Summer Institute 99
held at Fuji-Yoshida, the discussions at which motivated this work. 
T.~K.\ is supported in
part by the Grant-in-Aid for Scientific Research No.~10640261 from 
Japan Society for the Promotion of Science and 
the Grant-in-Aid for Scientific Research on Priority Areas No.~12047214
from the Ministry of Education, Science, Sports and Culture, Japan.

\appendix
\section{Conventions and Useful Identities}

Throughout this paper we use $\eta_{ab}={\rm diag}(+,-,-,-,-)$ 
as the Lorentz metric, which is different from that of BSVP but is 
more familiar to phenomenologists. With this metric, the 
$4\times4$ Dirac $\gamma$-matrices $\gamma^a$ in 5D satisfy, as usual,
\begin{equation}
\gamma^a\gamma^b+\gamma^b\gamma^a=2\eta^{ab},\qquad (\gamma^a)^\dagger=\eta_{ab}\gamma^b.
\end{equation}
We use the multi-index $\gamma$-matrices defined by
\begin{eqnarray}
\gamma^{(n)}\equiv\gamma^{a_1\cdots a_n}\equiv\gamma^{[a_1}\gamma^{a_2}\cdots\gamma^{a_n]}
\equiv{1\over n!}\sum_{p=\hbox{\scriptsize perm's}}
(-1)^p\gamma^{a_1}\gamma^{a_2}\cdots\gamma^{a_n},
\end{eqnarray}
where the square bracket $[\ \cdots\ ]$ attached to the indices implies complete 
antisymmetrization with {\it weight 1}. Similarly, $(\ \cdots\ )$ is used 
for complete symmetrization with weight 1. We choose the Dirac matrices
to satisfy
\begin{equation}
\gamma^{a_1\cdots a_5}=\epsilon^{a_1\cdots a_5},\qquad \epsilon^{01234}=1,
\end{equation}
where $\epsilon^{a_1\cdots a_5}$ is a totally antisymmetric tensor. 
With this choice, the duality relation reads 
\begin{equation}
  \gamma^{a_1\cdots a_n}={S_n\over(5-n)!}\epsilon^{a_1\cdots a_nb_1\cdots b_{5-n}}
      \gamma_{b_1\cdots b_{5-n}},\qquad 
  S_n=\left\{\begin{array}{cl} 
       +1&{\rm for}\quad  n=0,1,4,5\\
       -1&{\rm for}\quad  n=2,3       
      \end{array}\right. .
\end{equation}

The charge conjugation matrix $C$ in 5D has the properties 
\begin{equation}
C^\T=-C,\qquad C^\dagger C=1,\qquad C\gamma_aC^{-1}=\gamma_a^\T.
\end{equation}
Our spinors carry the {\it SU}$(2)$ index $i$ ($i=1,2$) of $\XU_{ij}$. 
This index is generally raised or lowered by using the $\epsilon$ symbol
$\epsilon^{ij}=\epsilon_{ij}=i(\sigma_2)_{ij}$:
\begin{equation}
\psi^i = \epsilon^{ij}\psi_j,\qquad  
\psi_i = \psi^j\epsilon_{ji}.
\end{equation}
[Note that contraction of the {\it SU}$(2)$ indices is always performed 
according to the northwest-to-southeast convention.] \ All our 5D spinors 
satisfy the {\it SU}$(2)$-Majorana condition,
\begin{eqnarray}
\bar\psi^i\equiv(\psi_i)^\dagger\gamma^0=(\psi^i)^\T C. 
\end{eqnarray}
When {\it SU}$(2)$ indices are suppressed in bilinear terms of spinors, 
the northwest-to-southeast contraction of the {\it SU}$(2)$ indices
is implied:
\begin{eqnarray}
\bar\psi\gamma^{(n)}\lambda\equiv\bar\psi^i\gamma^{(n)}\lambda_i.
\end{eqnarray}
The following symmetry of spinor bilinear terms is important:
\begin{equation}
\bar\lambda\gamma^{(n)}\psi=t_n\bar\psi\gamma^{(n)}\lambda,\qquad 
t_n=\left\{\begin{array}{cl} 
       +1&{\rm for}\quad  n=2,3\\
       -1&{\rm for}\quad  n=0,1,4,5
      \end{array}\right. .
\end{equation}
If the {\it SU}$(2)$ indices are not contracted, the sign becomes 
opposite.

We often use the Fierz identity, which in 5D reads
\begin{eqnarray}
\psi^i\bar\lambda^j=-\myfrac14(\bar\lambda^j\psi^i)-\myfrac14(\bar\lambda^j\gamma^a\psi^i)\gamma_a
+\myfrac18(\bar\lambda^j\gamma^{ab}\psi^i)\gamma_{ab}.
\end{eqnarray}
An important identity can be proved by 
using this Fierz identity: the antisymmetric trilinear in spinors 
$\psi_{1}$, $\psi_{2}$, $\psi_{3}$ satisfies
\begin{equation}
\gamma^a\psi^i_{[1}(\bar\psi^{\jspan}_2\gamma_a\psi^\ispan_{3]})
-\psi^i_{[1}(\bar\psi^\jspan_2\psi^\ispan_{3]})=0.
\end{equation}
From this identity and with the help of the Fierz identity again, when 
necessary, various identities can be derived and used when proving, e.g.,
the invariance of the action. 
Two examples are
\begin{eqnarray}
&&2\psi_{aj}(\bar\psi^i_b\gamma^{abc}\psi_c^j) -
3\gamma^{ab}\psi^i_{[a}(\bar\psi^\jspan_b\gamma^c\psi^\ispan_{c]})=0,  \nn
&&\bar\psi^\ispan_{[a}\gamma^d\psi^\jspan_b\bar\psi^{\jspan i}_c\gamma^{abc}\psi_{d]}^j=0.
\end{eqnarray}
(Six-dimensional analogues to these can actually be 
found more easily, from which one could derive these by 
dimensional reduction.) \ 

A useful formula in treating the {\it SU}$(2)$ indices is
\begin{equation}
A^{ij} = -\half\epsilon^{ij}A +A^{(ij)}, \quad 
A^i{}_j = +\half\delta^i_jA +A_S{}^i{}_j, 
\end{equation}
where $A\equiv A^i{}_i$ and $A_S{}^i{}_j\equiv A^{(ik)}
\epsilon_{kj}$. 
Often used is the following formula for $\gamma^{\mu_1\mu_2\cdots\mu_n}$, 
which is valid for any number of spacetime dimensions $d$: 
\begin{eqnarray}
&&\gamma^{\nu_1\nu_2\cdots\nu_m}
\gamma^{\mu_1\mu_2\cdots\mu_n}\gamma_{\nu_1\nu_2\cdots\nu_m}
=C^d_{n,m}\gamma^{\mu_1\mu_2\cdots\mu_n},\nn
&&C_{n,m}^d = (-1)^{[d/2]+nm}m!\times\hbox{\{%
Coefficient of $x^m$ in $(1+x)^{d-n}(1-x)^n$\}}.
\hspace{2em}
\label{eq:Case}
\end{eqnarray}
More explicit values for $C^d_{n,m}$ in $d=5$ and 6 cases 
are given in Table \ref{table:Case}.

Finally, we mention the formulas holding for $d=6$ for the (anti-)self-dual 
tensor $T^\pm_{abc}= \pm(1/6)\epsilon_{abcdef}T^{\pm\,def}$, which are 
useful in the dimensional reduction: 
\begin{eqnarray}
\gamma\dt T^{\pm}&=&6T^{\pm}_{ab5}\gamma^{ab5}{\cal P}^\pm,\qquad 
\left({\cal P}^\pm={1\pm\gamma_7\over2}\right), \nn
\gamma_a\gamma\dt T^{\pm}&=&6T^{\pm}_{abc}\gamma^{bc}{\cal P}^{\pm},\qquad 
\gamma_{[a}\gamma\dt T^{\pm}\gamma_{b]}=-12T^{\pm}_{abc}\gamma^c{\cal P}^{\mp}.
\end{eqnarray}
\begin{table}[bt]
\caption{The coefficients $C^d_{n,m}$ in the formula 
Eq.~(\protect\ref{eq:Case}) 
for $d=5$ and 6 cases.}
\label{table:Case}
\parbox{5.8cm}{
\begin{flushright}
\begin{tabular}{l|rr} \hline\hline
$d=5$     & $m=1$ & $m=2$  \\ \hline
$n=0$     & $5$ & $-20$  \\
$n=1$     & $-3$ & $-4$  \\
$n=2$     & $1$  & $4$  \\ 
$n=3$     & $1$  & $4$  \\ \hline
\end{tabular}
\end{flushright}
}
\hspace{1cm}
\parbox{\halftext}{
\begin{tabular}{l|rrr} \hline\hline
$d=6$        & $m=1$ & $m=2$ & $m=3$ \\ \hline
$n=0$     & $6$  & $-30$  & $-120$ \\
$n=1$     & $-4$ & $-10$  & $0$ \\
$n=2$     & $2$  & $2$   & $24$ \\ 
$n=3$     & $0$  & $6$   & $0$ \\ \hline
\end{tabular}}
\end{table}

\section{6D Superconformal Tensor Calculus and Correspondence}

Since our notation and, in particular, metric convention are different 
from BSVP, we here summarize the results of BSVP for 6D superconformal 
tensor calculus for the parts we need in this paper using our notation. 
For reader's convenience, we also give the correspondence between the 
BSVP notation and our notation in Table \ref{table:D}. In this appendix,
we omit the underlines that indicate six dimensional quantities. 
\begin{table}[bt]
\caption{Correspondence between the notation of BSVP and that used here.}
\label{table:D}
\def\ssquare{{\mathchoice{\sqr34}{\sqr34}{\sqr{2.1}3}{\sqr{1.5}3}}}
\begin{center}
\begin{tabular}{c|c} \hline\hline
BSVP &  Ours \\ \hline
{$x_\mu$}          & $x^\mu$ \ or \ $-x_\mu$ \\ 
{$\partial_\mu$}         & $\partial_\mu$ \ or \ $-\partial^\mu$ \\ 
{$\delta_{ab}$}      & $-\eta_{ab}$  \\ 
{$\gamma_a$}         & $-i\gamma^a$ \ or \ $i\gamma_a$ \\ 
{$\gamma^a\partial_a\equiv\slash{\partial}$} & $-i\gamma^a\partial_a=-i\slash{\partial}$  \\ 
{$\gamma_7=i\gamma_1\gamma_2\gamma_3\gamma_4\gamma_5\gamma_6$}         
& $\gamma_7= \gamma_0\gamma_1\gamma_2\gamma_3\gamma_4\gamma_5$ \\
\rule[-1.15ex]{0pt}{1ex}
{$\epsilon_{abcdef}$}\quad ($\epsilon_{123456}=+1$)    
  & $-i\epsilon_{abcdef}$ \ or \ $-i\epsilon^{abcdef}$ 
\quad ($\epsilon^{012345}=+1$)\\ \hline
\multicolumn{2}{c}{Superconformal generators $\XX_A$} \\ \hline
$\XP_a, \ \ \XQ, \ \ \XD, \ \ \XK_a $ & 
$\XP_a,\ \ (1/2)\XQ,\ \ \XD, \ \ -\XK_a$ \\ 
$\XM_{ab},\ \ \XU, \ \ \XS$ &
 $\XM_{ab},\ \ (1/2)\XU,\ \ (1/2)\XS$ \\ \hline
\multicolumn{2}{c}{
\rule[-1.1ex]{0pt}{1ex}
\Tspan{Gauge fields $h_\mu^A$ (the same rules for 
$R_{\mu\nu}^A$ and transformation parameters $\varepsilon^A$)}} \\ \hline
\Tspan{$e_\mu^a,\ \ \psi_\mu,\ \ b_\mu,\ \ f_\mu^a$} &
$e_\mu^a,\ \ {2}\psi_\mu,\ \ b_\mu,\ \ -f_\mu^a$ \\
$\omega_\mu^{ab},\ \ V_\mu^{ij},\ \ \phi_\mu$ & 
$\omega_\mu^{ab},\ \ {2}V_\mu^{ij},\ \ {2}\phi_\mu$ \\
\rule[-1.15ex]{0pt}{1ex}
$T^-_{abc}$ &   $T^-_{abc}$ \ or \ $-T^{-\,abc}$  \\
\hline
\end{tabular}
\end{center}
\end{table}

The local superconformal tensor calculus is constructed generally by 
deforming the Yang-Mills theory based on the superconformal
group.\cite{ref:KTN,ref:vN,ref:dWvHVP,ref:VP} \  
In 6D, the superconformal group is $OSp(6,2|1)$, whose generators are given 
in Eq.~(\ref{eq:6Dgener}), and the algebra is deformed by imposing 
the following three constraints on the curvatures:
\begin{eqnarray}
  &&\hat{R}_{\mu\nu}{}^a(P)  =  0\,,\nn
  &&\hat{R}_{\mu\nu}{}^{ba}(M)e^\nu{}_b-T^-_{\mu bc}T^{-abc}
    -{1\over12}e_\mu{}^aD= 0\,,\nn
  &&\gamma^\mu\hat{R}_{\mu\nu}{}^i(Q) ={1\over12}\gamma_\nu\chi^i\,.
  \label{eq:6D.cons}
\end{eqnarray}
Note that the $\hat{R}_{\mu\nu}(\XX)$ are the covariant curvatures of the 
resultant algebra, but not of the original superconformal group.

\subsection{Weyl multiplet}

As mentioned in Eq.~(\ref{eq:WeylM}), the $d=6$ Weyl multiplet 
consists of 40 Bose plus 40 Fermi fields, whose properties are summarized 
in Table \ref{table:A1}. 
The $\XM$, $\XK$ and $\XS$ gauge fields 
$\omega_{\mu}{}^{ab},\ f_{\mu}^{a}$ and $\phi_{\mu}^i$ become 
dependent fields by the constraints (\ref{eq:6D.cons}). 
For example, the $\omega_{\mu}{}^{ab}$ 
is given as follows 
by solving the first constraint $\hat{R}_{\mu\nu}{}^a(P)  =  0$:
\begin{eqnarray}
  \omega_\mu^{\ ab}& = &
   \omega_\mu^{0\ ab}+i(2\bar\psi_\mu\gamma^{[a}\psi^{b]} 
    +\bar\psi^a\gamma_\mu\psi^b)-2e_\mu^{\ [a}b^{b]}\,, \nn
   \omega_\mu^{0\ ab}&\equiv&-2e^{\nu[a}\partial_{[\mu}e_{\nu]}^{\ b]}
    +e^{\rho[a}e^{b]\sigma}e_\mu{}^c\partial_\rho e_{\sigma c}\,.
\label{eq:6Dspinconn}
\end{eqnarray}

\begin{table}[tb]
\caption{Weyl multiplet.}
\label{table:A1}
\begin{center}
\begin{tabular}{ccccc} \hline \hline
    field      & type   & restrictions & {\it SU}$(2)$ & Weyl-weight    \\ \hline 
\Tspan{$e_\mu{}^a$} &   boson    & sechsbein    & \bf{1}    &  $ -1$     \\  
$\psi^i_\mu$  &  fermion  & $\gamma_7\psi^i_\mu=+\psi^i_\mu$ & \bf{2}  &$-\hspace{-2mm}\myfrac12$ \\  
\Tspan{$V^{ij}_\mu$}    &  boson    & $V_\mu^{ij}=V_\mu^{ji}=-V^*_{\mu ij}$   & \bf{3}&0\\ 
$b_\mu$ & boson & real & \bf{1} & 0\\
$T^-_{abc}$&boson&$T^-_{abc}=-\myfrac16\epsilon_{abcdef}T^{-def}$&\bf{1}&1 \\
$\chi^i$  &  fermion  & $\gamma_7\chi^i=+\chi^i$ & \bf{2}    &\hspace{-1.5mm}\myfrac32 \\  
$D$    &  boson    & real & \bf{1} & 2 \\ \hline
\multicolumn{5}{c}{dependent gauge fields} \\ \hline
\Tspan{$\omega_\mu{}^{ab}$} &   boson    & spin connection & \bf{1}    &   0     \\  
$\phi^i_\mu$  &  fermion  & $\gamma_7\phi^i_\mu=-\phi^i_\mu$ & \bf{2}  &$\hspace{-1.5mm}\myfrac12$ \\  
$f_\mu{}^a$ &  boson    & real & \bf{1}&1\\ \hline
\end{tabular}
\end{center}
\end{table}

The gauge fields of the Weyl multiplet transform under the $\XQ,\XS$ and 
$\XK$ transformation 
$\delta\equiv\bar{\varepsilon}^i\XQ_i+\bar{\eta}^i\XS_i+\xi_K^a\XK_a\equiv 
\delta_Q({\varepsilon})+\delta_S(\eta)+\delta_K(\xi_K)$
as
  \begin{eqnarray} 
  \delta e_\mu^a& = &-2i\bar{\varepsilon}\gamma^a\psi_{\mu} \,,\nn 
  \delta\psi_\mu^i& = &{\cal D}_\mu\varepsilon^i
      -\myfrac{1}{24}\,\gamma\dt T^-\gamma_\mu\varepsilon^i + i\gamma_\mu\eta^i\,,\nn
  \delta\omega_\mu{}^{ab}& = &
    +2\bar{\varepsilon}\gamma^{ab}\phi_\mu 
    -2i\bar{\varepsilon}\gamma^{[a}\hat{R}_\mu{}^{b]}(Q)
    -i\bar{\varepsilon}\gamma_\mu\hat{R}^{ab}(Q) \nn &&{}
  +\myfrac{i}{6}e_\mu{}^{[a}\bar{\varepsilon}\gamma^{b]}\chi 
  -2i\bar{\varepsilon}\gamma_c\psi_\mu T^{-abc} 
   +2\bar{\eta}\gamma^{ab}\psi_\mu 
      -4\xi_K^{[a}e_\mu{}^{b]}\,,\nn 
  \delta b_\mu& =
   &-2\bar{\varepsilon}\phi_\mu 
    -\myfrac{i}{12}\bar{\varepsilon}\gamma_\mu\chi+2\bar{\eta}\psi_\mu-2\xi_K^ae_{\mu a}\,, \nn 
  \delta V_\mu{}^{ij}& = &-8\bar{\varepsilon}^{(i}\phi_\mu^{j)}
    -\myfrac{i}{3}\bar{\varepsilon}^{(i}\gamma_\mu\chi^{j)} -8\bar{\eta}^{(i}\psi_\mu^{j)}\,,\nn
  \delta\phi_\mu^i& = &if_\mu{}^a\gamma_a\varepsilon^i
         -\myfrac{i}{16}(\gamma^{ab}\gamma_\mu 
    -\myfrac{1}{2}\gamma_\mu\gamma^{ab})\hat{R}_{ab}{}^i{}_j(U)\varepsilon^j \nn
    &&{}-\myfrac{i}{96}(\slash{\hat{\cal D}}\,\gamma\dt T^-)\gamma_\mu\varepsilon^i
    +\myfrac{1}{12}\gamma_a\chi^i\,\bar{\varepsilon}\gamma^a\psi_\mu 
    +{\cal D}_\mu\eta^i -i\xi_K^a\gamma_a\psi_\mu^i \,, 
 \label{eq:6DWeylTrf}
  \end{eqnarray}
where the derivative ${\cal D}_\mu$ is covariant only with respect 
to the homogeneous transformations $\XM_{ab},\,\XD$ and $\XU_{ij}$ 
(and $\XG$ for non-singlet fields under the Yang-Mills group $G$).
These homogeneous (i.e., Weyl weight $w=0$) transformations 
$\XM_{ab},\,\XD$ and $\XU_{ij}$ are self-evident from the index structure 
carried by the fields; our conventions are
\begin{eqnarray}
&&\delta_M(\lambda)\phi^a=\lambda^a{}_b\phi^b,\qquad
\qquad \delta_M(\lambda)\psi=\myfrac14\lambda^{ab}\gamma_{ab}\psi\ \ (\psi: \hbox{spinor}), \nn
&&\delta_U(\theta)\phi^i=\theta^i{}_j\phi^j,\ \ \quad\,\rightarrow\ \quad\,
\delta_U(\theta)\phi_i=\theta_{ij}\phi^j=-\theta_i{}^j\phi_j, \nn
&&\delta_G(t)\phi^\alpha=t^\alpha{}_\beta\phi^\beta,\, \qquad \qquad
\delta_D(\rho)\phi=w\rho\phi\ \ (w: \hbox{Weyl weight of } \phi).
\label{eq:homotrfs}
\end{eqnarray}

The $\XQ,\,\XS,\,\XK$ transformation law of the `matter multiplet' is
  \begin{eqnarray} 
  \delta T^-_{abc} & = &
  -\myfrac{i}{8}\bar\varepsilon\gamma^{de}\gamma_{abc}\hat{R}_{de}(Q)
   +\myfrac{7}{48}i\,\bar{\varepsilon}\gamma_{abc}\chi\,,\nn 
  \delta\chi^i & = &
   \myfrac{1}{4}(\hat{\calD}_\mu\,\gamma\dt T^-)\gamma^\mu\varepsilon^i
   +\myfrac{3}{4}\,\gamma\dt {\hat R}^i{}_j(U)\varepsilon^j
   +\myfrac{1}{2}D\varepsilon^i+i\gamma\dt T^-\eta^i\,,\nn 
  \delta D & = & -2i\bar{\varepsilon}\slash{\hatD}\chi-4\bar{\eta}\chi\,.
\label{eq:6DMTrf}
 \end{eqnarray}
The algebra (\ref{eq:Galg}) of these transformations is 
given explicitly by
  \begin{eqnarray}
  {}[\delta_Q(\varepsilon_1),\delta_Q(\varepsilon_2)] & = &
\xi^\mu\hat\calD_\mu 
+\delta_M(\xi_cT^{-abc})
+\delta_S\bigl(\myfrac{i}{24}\xi^\mu\gamma_\mu\chi^i\bigr) \nn
&&{}-\delta_K\bigl(\myfrac14\xi_b\hat\calD_cT^{-abc}
+\myfrac{1}{48}\xi^aD\bigr)\,, 
\qquad 
\bigl(\ \xi^a \equiv2i(\bar\varepsilon_1\gamma^a\varepsilon_2)\ \bigr) \nn 
  {}[\delta_S(\eta),\delta_Q(\varepsilon)] & = &
    \delta_D(-2\bar\varepsilon\eta)+\delta_M(2\bar{\varepsilon}\gamma^{ab}\eta)
     +\delta_U(-8\bar{\varepsilon}^{(i}\eta^{j)})\,,\nn
  {}[\delta_S(\eta_1),\delta_S(\eta_2)] & = &
    \delta_K(2i\bar{\eta}_1\gamma^a\eta_2)\,.
\label{eq:6D.algebra}
  \end{eqnarray} 
This algebra can be read directly from the structure
functions which appear in the transformation laws (\ref{eq:6DWeylTrf}) 
of the Weyl multiplet gauge fields (cf.\ Eq.~(\ref{eq:gaugeTrf})). 

\subsection{Vector multiplet}

\begin{table}[tb]
\caption{Matter multiplets.}
\label{table:A2}
\begin{center}
\begin{tabular}{ccccc}  \hline \hline
  fields      & type    & restrictions   & {\it SU}$(2)$&Weyl-weight    \\ \hline 
\multicolumn{5}{c}{Vector multiplet} \\ \hline
$W_\mu$      &  boson    & Yang-Mills gauge boson&  \bf{1}    &   0     \\  
$\Omega^i$      &  fermion  & $\gamma_7\Omega^i=+\Omega^i$           & \bf{2} &\hspace{-1.5mm}\myfrac32 \\  
$Y_{ij}$    &  boson    & $Y^{ij}=Y^{ji}=-Y^*_{ij}$   & \bf{3} & 2 \\ \hline
\multicolumn{5}{c}{Linear multiplet} \\ \hline
\Tspan{$L^{ij}$} & boson & $L^{ij}=L^{ji}=-(L_{ij})^*$ & \bf{3} & 4 \\ 
$\varphi^i$ & fermion & $\gamma_7\varphi^i=-\varphi^i$ & \bf{2} &\hspace{-1.5mm}\myfrac92\\
$E_a$    &  boson    & real & \bf{1} & 5 \\ \hline
\multicolumn{5}{c}{Nonlinear multiplet} \\ \hline
\Tspan{$\Phi^i_\alpha$}  &  boson    & $(\Phi^\alpha_i)^*=-\Phi^i_\alpha,\quad 
\Phi^i_\alpha\Phi^\alpha_i=\delta^i_j,\quad \Phi^\alpha_i\Phi^i_\beta=\delta^\alpha_\beta$  &  \bf{2}   & 0\\ 
$\lambda^i$   &  fermion  & $\gamma_7\lambda^i=-\lambda^i$           &
\bf{2}    &\hspace{-1.5mm}\myfrac12\\ 
$V_a$    &  boson    & real & \bf{1} & 1 \\ \hline
\multicolumn{5}{c}{Hypermultiplet} \\ \hline
\Tspan{$\hA^i_\alpha$}     &  boson    
   & $\hA^i_\alpha=\epsilon^{ij}\hA_j^\beta\rho_{\beta\alpha}=-(\hA_i^\alpha)^*$   & \bf{2}  & 2  \\  
$\zeta_\alpha$    &  fermion  & $\gamma_7\zeta_\alpha=-\zeta_\alpha$    &
\bf{1}  &\hspace{-1.5mm}\myfrac52\rule[-3mm]{0pt}{1mm}\\ \hline
\end{tabular}
\end{center}
\end{table}
The vector multiplet components are summarized in Table \ref{table:A2}
and they are all Lie-algebra valued; 
e.g., $W_\mu$ represents the matrix 
$W_\mu{}^\alpha{}_\beta=W_\mu^A(t_A)^\alpha{}_\beta$, where the $t_A$ 
are the (anti-hermitian) 
generators of the gauge group $G$.
The $\XQ,\,\XS,\,\XK$ gauge transformation law is 
\begin{eqnarray}
   \delta W_\mu& = &-2i\bar{\varepsilon}\gamma_\mu\Omega\,, \nn
   \delta\Omega^i& = &-\myfrac14\,\gamma\dt\hat F(W)\varepsilon^i-Y^{ij}\varepsilon_j\,, \nn
   \delta Y^{ij}& = &2i\bar\varepsilon^{(i}\slashD \Omega^{j)}+4\bar\eta^{(i}\Omega^{j)}\,,
\label{eq:6DVtrf}
\end{eqnarray}
where the covariant field strength is 
$\hat F_{\mu\nu}(W)\equiv F_{\mu\nu}(W)+4i\bar\psi_{[\mu}\gamma_{\nu]}\Omega$, with
$F_{\mu\nu}(W)\equiv2\partial_{[\mu}W_{\nu]}-g[W_\mu,\,W_\nu]$, and 
$\hatD_\mu$ is covariant also with respect to $\XG$.
\subsection{Linear multiplet}
The linear multiplet consists of the components given in 
Table \ref{table:A2}, and they may generally carry non-Abelian charge.
Their $\XQ,\,\XS,\,\XK$ transformation rules are
\begin{eqnarray}
\delta L^{ij}&=&2\bar\varepsilon^{(i}\varphi^{j)}\,,\nn
\delta\varphi^i&=&-i\slashD L^{ij}\varepsilon_j
+\myfrac{i}2\gamma^a\varepsilon^iE_a-8L^{ij}\eta_j\,,\nn
\delta E_a&=&2\bar\varepsilon\gamma_{ab}\hatD^b\varphi-\myfrac1{12}\bar\varepsilon\gamma_a\gamma\dt
T^-\varphi-\myfrac23i\bar\varepsilon^i\gamma_a\chi^jL_{ij}\nn
&&{}-4ig\bar\varepsilon_i\gamma_a\Omega_jL^{ij} -10i\bar\eta\gamma_a\varphi \,.
\label{eq:6DLtrf}
\end{eqnarray}
Closure of the algebra requires the following $\XQ$- and $\XS$-invariant 
constraint:
\begin{equation}
\hatD^aE_a+\myfrac12\bar\varphi\chi+4g\bar\Omega\varphi+2gY^{ij}L_{ij}=0\,.
\label{eq:6DLcons}
\end{equation}

\subsection{Nonlinear multiplet}
The nonlinear multiplet consists of the components 
listed in Table \ref{table:A2}. 
Their $\XQ,\,\XS,\,\XK$ transformation
rules are
\begin{eqnarray}
\delta\Phi^i_\alpha&=&2\bar\varepsilon^{(i}\lambda^{j)}\Phi_{j\alpha}\,,\nn
\delta\lambda^i&=&-i\Phi^i_\alpha\slashD \Phi^\alpha_j\varepsilon^j+\myfrac{i}2\gamma^aV_a\varepsilon^i
+\myfrac12\gamma_a\varepsilon_j\,\bar\lambda^i\gamma^a\lambda^j \nn
&&{}+\myfrac1{48}\gamma_{abc}\varepsilon^i\,\bar\lambda\gamma^{abc}\lambda-4\eta^i\,,\nn
\delta V_a&=&2\bar\varepsilon\gamma_{ab}\hatD^b\lambda+\myfrac{i}3\bar\varepsilon\gamma_a\chi 
-\bar\varepsilon\gamma_a\gamma^b\lambda V_b-\myfrac1{12}\bar\varepsilon\gamma_a\gamma\dt T^-\lambda\nn
&&{}-2\bar\varepsilon_i\gamma_a\Phi^i_\alpha\slashD \Phi^\alpha_j\lambda^j-4ig\bar\varepsilon^i\gamma_a\Omega^\alpha_{j\beta}
\Phi^j_\alpha\Phi^\beta_i-2i\bar\eta\gamma_a\lambda-8\xi_{Ka}\,.
\hspace{2em}
\label{eq:6DNLtrf}
\end{eqnarray}
(We stick to the group transformation convention in (\ref{eq:homotrfs}), 
so that the sign of $g$ here is opposite to that of BSVP.)
Closure of the algebra requires the following $\XQ$-, $\XS$-, 
$\XK$-invariant constraint:
\begin{eqnarray}
\hatD^aV_a+\myfrac13D-\myfrac12V^aV_a+\hatD^a\Phi^i_\alpha\hatD_a\Phi^\alpha_
i+2i\bar\lambda\slashD\lambda-\myfrac56\bar\lambda\chi-\myfrac{i}6\bar\lambda\,\gamma\dt T^-\lambda\nn
{}+2i\bar\lambda^i\Phi_{\alpha i}\slashD \Phi^\alpha_j\lambda^j+2gY_{ij}{}^\alpha_\beta\Phi^i_\alpha\Phi^{\beta 
j}+8g\bar\lambda^i\Omega_j{}^\alpha_\beta\Phi^j_\alpha\Phi^\beta_i=0\,.
\hspace{2em}
\end{eqnarray}

\subsection{Hypermultiplet}

The hypermultiplet in 6D is an on-shell multiplet consisting of 
scalars $\hA^i_\alpha$ and negative chiral spinors $\zeta_\alpha$. They carry 
the index $\alpha\ (=1,2,\cdots,2r)$ of the gauge group $G$, which is raised 
or lowered by using a $G$-invariant tensor $\rho_{\alpha\beta}$ (and $\rho^{\alpha\beta}$ 
with $\rho^{\gamma\alpha}\rho_{\gamma\beta}=\delta^\alpha_\beta$) like 
$\hA^i_\alpha=\hA^{i\beta}\rho_{\beta\alpha}$. The
tensor $\rho_{\alpha\beta}$ can generally be brought into the 
standard form
\begin{equation}
\rho_{\alpha\beta}=\pmatrix{
    \epsilon    &       &      \cr
        &\epsilon       &      \cr
        &       & \ddots        \cr} = \rho^{\alpha\beta},
\qquad
\epsilon= \pmatrix{ 0 & 1 \cr -1 & 0 \cr}.
\label{eq:stdrho}
\end{equation}
The scalar field $\hA^i_\alpha$ also carries the 
superconformal {\it SU}$(2)$ index $i$ and satisfies the reality condition
\begin{equation}
\hA^i_\alpha=\epsilon^{ij}\hA_j^\beta\rho_{\beta\alpha}=-(\hA_i^\alpha)^*,\qquad
\hA_{i\alpha}=(\hA^{i\alpha})^*,
\label{eq:Areal}
\end{equation}
so that $\hA^i_\alpha$ can be identified with $r$ quaternions 
${\mbf q}_l\equiv q^0_l+{\mbf i}q^1_l+{\mbf j}q^2_l+{\mbf k}q^3_l$ 
$(l=1,\cdots,r)$; indeed, by the above condition, the $l$-th $2\times2$ matrix 
$(\hA^i_{\alpha=2l-1},\,\hA^i_{\alpha=2l})$ can be written in the form 
$q^0_l{\mbf1}_2+iq^1_l\sigma_1+iq^2_l\sigma_2+iq^3_l\sigma_3$ with Pauli matrices 
$\sigma_1$, $\sigma_2$ and $\sigma_3$. 
Thus the group $G$ acting on the the hypermultiplet 
should be a subgroup of {\it GL}$(r; {\mbf H})$:
\begin{eqnarray}
&&\delta_G(t)\hA^\alpha_i=gt^\alpha{}_\beta\hA^\beta_i, \qquad  
\delta_G(t)\hA^i_\alpha=g(t^\alpha{}_\beta)^*\hA^i_\beta=-gt_\alpha{}^\beta\hA_\beta^i, \nn
&&t_\alpha{}^\beta\equiv\rho_{\alpha\gamma}t^\gamma{}_\delta\rho^{\delta\beta}=-(t^\alpha{}_\beta)^*.
\label{eq:H.Gtr}
\end{eqnarray}

The $\XQ$ and $\XS$ transformations of the hypermultiplet are given by
\begin{equation}
\delta\hA^i_\alpha= 2\bar\varepsilon^i\zeta_\alpha, \qquad 
\delta\zeta^\alpha= i\slashD \hA^\alpha_j\varepsilon^j + 4\hA_j^\alpha\eta^j.
\label{eq:6DhyperTrf}
\end{equation}
With these rules, the superconformal algebra
(\ref{eq:6D.algebra}) is not realized on $\zeta_\alpha$:
\begin{eqnarray}
{}[\delta_Q(\varepsilon_1),\ \delta_Q(\varepsilon_2)] \zeta^\alpha= 
[\hbox{RHS of Eq.~(\ref{eq:6D.algebra})}]\zeta^\alpha 
-\gamma_a\Gamma^\alpha\,(\bar\varepsilon_1\gamma^a\varepsilon_2) .
\label{eq:6D.uncl}
\end{eqnarray}
Therefore, $\Gamma^\alpha=0$ should be an equation of motion for $\zeta^\alpha$ for 
the algebra to close on-shell. 
This fermionic non-closure function $\Gamma^\alpha$ closes under the superconformal 
transformations together with its bosonic partner 
$C^\alpha_i$:
\begin{eqnarray}
\Gamma^\alpha&\equiv& -i\slashD\zeta^\alpha-\myfrac13\hA_j^\alpha\chi^j+
\myfrac{i}{12}\gamma\cdot T^-\zeta^\alpha-2g\Omega^{i\,\alpha}_{\ \ \,\beta}\hA^\beta_i,\nn
C^\alpha_i&\equiv& (-\hatD^a\hatD_a+\myfrac16D)\hA_i^\alpha+\myfrac16\bar\zeta^\alpha\chi_i
+4g\bar\Omega_{i\ \,\beta}^{\ \alpha}\zeta^\beta-2gY_{ij\ \beta}^{\ \,\alpha}\hA^{\beta j}, 
\label{eq:defGammaC} \\
\delta\Gamma^\alpha&=& -C^\alpha_i\varepsilon^i +i\gamma^\mu\gamma^a\Gamma^\alpha\,(\bar\varepsilon\gamma_a\psi_\mu), \nn
\delta C^\alpha_i&=& -2i\bar\varepsilon_i\slashD\Gamma^\alpha 
-2(\bar\psi_{\mu i}\gamma^a\psi^\mu_j)(\bar\varepsilon^j\gamma_a\Gamma^\alpha)
+4\bar\eta_i\Gamma^\alpha.
\label{eq:6D.uncl.trs}
\end{eqnarray}
Here, since the algebra does not close on $\Gamma^\alpha$ either, 
the covariant derivative $\hatD_\mu$ acting on $\Gamma^\alpha$ is defined, 
slightly differently 
from the usual definition (\ref{eq:covD}), to be
\begin{equation}
\hatD_\mu\Gamma^\alpha=\calD_\mu\Gamma^\alpha+C_i^\alpha\psi_\mu^i
-\half i\gamma^\nu\gamma^a\Gamma^\alpha(\bar\psi_\mu\gamma_a\psi_\nu),
\label{eq:6D.uncl.covD}
\end{equation}
so that $\hatD_\mu\Gamma^\alpha$ becomes covariant.

\subsection{Invariant action formulas}

The action for the product of a vector multiplet and a
linear multiplet,
\begin{eqnarray}
e^{-1}{\cal L}_{\rm VL}&=&Y^{ij}L_{ij}+2\bar\Omega\varphi
+2i\bar\psi^a_i\gamma_a\Omega_jL^{ij}\nn
&&{}-\myfrac12W_a(E^a-2\bar\psi_b\gamma^{ba}\varphi
+2i\bar\psi_b^i\gamma^{abc}\psi_c^jL_{ij})\,, 
\label{eq:VLaction}
\end{eqnarray}
is superconformal invariant if the vector multiplet is abelian and the
linear multiplet carries no gauge group charges or is charged only
under the abelian group of this vector multiplet. 

The invariant action for the hypermultiplet, which gives as the 
equations of motion the required on-shell closure condition $\Gamma_\alpha=0$ as
well as its bosonic partner $C_\alpha^i=0$, is given by
\begin{equation}
e^{-1}\calL =  
\hA_i^\alpha d_\alpha{}^\beta C_\beta^i
+2(\bar\zeta^\alpha-i\bar\psi^i_\mu\gamma^\mu\hA_i^\alpha)d_\alpha{}^\beta\Gamma_\beta\,.
\label{eq:6D.Act.H}
\end{equation}
Here $d_\alpha{}^\beta$ is a $G$-invariant hermitian tensor satisfying
\begin{eqnarray}
&&(d_\alpha{}^\beta)^*=d_\beta{}^\alpha, \qquad 
d_{\alpha\beta}=-d_{\beta\alpha}, \quad (d_{\alpha\beta}\equiv d_\alpha{}^\gamma\rho_{\gamma\beta}) \nn
&&t^\gamma{}_\alpha d_\gamma{}^\beta+d_\alpha{}^\gamma(t^\gamma{}_\beta)^*=0.
\label{eq:tdrel}
\end{eqnarray}
It is shown in Ref.~\citen{ref:dWLVP} that field redefinitions can
simultaneously bring $\rho_{\alpha\beta}$ into the standard form
(\ref{eq:stdrho}) and $d_\alpha{}^\beta$ into the form
\begin{equation}
d_\alpha{}^\beta=\pmatrix{{\bf{1}}_p&  \cr
        &-{\bf{1}}_q \cr}.
\qquad (p,q: \hbox{even})
\label{eq:dstf}
\end{equation}
This property and the condition (\ref{eq:H.Gtr}) imply that the 
gauge group $G$ should be a subgroup of {\it USp}$(p,q)$.




\begin{thebibliography}{99}
\bibitem{hierarchy} 
  I.\ Antoniadis, \PL{246B,1990,377}. \\
  J.~D.\ Lykken, \PR{D54,1996,3693}. \\
  N.\ Arkani-Hamed, S.\ Dimopoulos and G.\ Dvali,
  \PL{429B,1998,263}. \\
  I.\ Antoniadis, N.\ Arkani-Hamed, 
  S.\ Dimopoulos and G.\ Dvali, \PL{436B,1998,257}. 

\bibitem{susy} 
  I.\ Antoniadis, C.\ Mu\~noz and M.\ Quir\'os, 
  \NP{B397,1993,515}. \\
  I.\ Antoniadis, S.\ Dimopoulos, A.\ Pomarol and M.\ Quiros, 
  \NP{B544,1999,503}.

\bibitem{ref:MP} 
E.~A.\ Mirabelli and M.~E.\ Peskin,
\PR{D58,1998,065002}, {\tt hep-th/9712214}.

\bibitem{LutySund} 
L.\ Randall and R.\ Sundrum, \NP{B557,1999,79}. \\
M.~A.\ Luty and R.\ Sundrum, {\tt hep-th/9910202}.

\bibitem{fermion} 
  K.~R.\ Dienes, E.\ Dudas and T.\ Gherghetta, 
 \PL{436B,1998,55}; \NP{B537,1999,47}. \\
  S.\ Abel and S.\ King, \PR{D59,1999,095010}. \\
  N.\ Arkani-Hamed and S.\ Dimopoulos, {\tt hep-ph/9811353}. \\
  N.\ Arkani-Hamed and M.\ Schmaltz, \PR{D61,2000,033005}, 
  {\tt hep-ph/9903417}.\\
  H.-C.\ Cheng, \PR{D60,1999,075015}.\\
  K.\ Yoshioka, Mod.~\PL{A15,2000,29}, {\tt hep-ph/9904433}.

\bibitem{astro} 
  N.\ Arkani-Hamed, S.\ Dimopoulos and G.\ Dvali, 
  \PR{D59,1999,086004}.\\
  M.\ Maggiore and A.\ Riotto, \NP{B548,1999,427}. \\
  N.\ Kaloper and A.\ Linde, \PR{D59,1999,101303}. \\
  N.\ Arkani-Hamed, S.\ Dimopoulos, N.\ Kaloper and J.\ March-Russell,
  \NP{B567,2000,189}, {\tt hep-ph/9903224}; {\tt hep-ph/9903239}. \\
  A.\ Riotto, \PR{D61,2000,123506}, {\tt hep-ph/9904485}.

\bibitem{ref:Pol} 
  J.~Polchinski, {\it String Theory, vol.~1, 2} (Cambridge
  Univ.~Press, Cambridge, 1998).

\bibitem{ref:HW} 
P.\ Ho\v{r}ava and E.\ Witten,
\NP{B460,1996,506}, {\tt hep-th/9510209}; 
\NP{B475,1996,94}, {\tt hep-th/9603142}.

\bibitem{ref:Zucker}
 M.~Zucker,
 \NP{B570,2000,267}, {\tt hep-th/9907082}; 
 {\tt hep-th/9909144}.

\bibitem{ref:CFGVP} 
E.\ Cremmer, S.\ Ferrara, L.\ Girardello and A.\ Van Proeyen, 
\PL{116B,1982,231}; \NP{B212,1983,413}.

\bibitem{ref:BSVP} 
E.\ Bergshoeff, E.\ Sezgin and A.\ Van Proeyen,
\NP{B264,1986,653}.

\bibitem{ref:KU2}
T.\ Kugo and S.\ Uehara,
\NP{B222,1983,125}.

\bibitem{ref:onoff} 
B.\ de Wit, J.~W.\ van Holten and A.\ Van Proeyen, 
\PL{95B,1980,51}, \\
P.\ Breitenlohner and M.~F.\ Sohnius,
\NP{B187,1981,409}. 

\bibitem{ref:CJS}
E.\ Cremmer, B.\ Julia and J.\ Scherk,
\PL{76B,1978,409}. \\
E.\ Cremmer and B.\ Julia,
\NP{B159,1979,141}. 

\bibitem{ref:Sohnius} 
M.~F.\ Sohnius, \JL{Z.\ Phys.,C18,1983,229}.

\bibitem{ref:VP} 
A.\ Van Proeyen, 
Lecture in the {\it Proceedings of the Winter School in Karpacz}, 1983, 
ed.\ B.\ Milewski (World Scientific Pub.\ Co.).

\bibitem{ref:dWLVP} 
B.\ de Wit, P.~G.\ Lauwers and A.\ Van Proeyen, 
\NP{B255,1985,569}.

\bibitem{ref:SuperspZ} 
M.~F.\ Sohnius, \NP{B138,1978,109}. \\
A.\ Galperin, E.\ Ivanov, S.\ Kalitzin, V.\ Ogievetsky and E.\ Sokatchev, 
\JL{Class.\ Quant.\ Grav.,1,1984,447}. 

\bibitem{ref:KO2} 
T.\ Kugo and K.\ Ohashi, in preparation.

\bibitem{ref:KU} 
T.~Kugo and S.~Uehara,
\NP{B226,1983,49}; \PTP{73,1985,235}. 

\bibitem{ref:dWvHVP} 
B.\ de Wit, J.~W.\ van Holten and A.\ Van Proeyen, 
\NP{B167,1980,186}; \andvol{B172,1980,543} (E); \ 
\andvol{B184,1981,77}; \andvol{B222,1983,516} (E). \\
M.\ de Roo, J.~W.\ van Holten, B.\ de Wit and A.\ Van Proeyen, 
\NP{B173,1980,175}. \\
E.\ Bergshoeff, M.\ de Roo and B.\ de Wit,
\NP{B182,1981,173}.

\bibitem{ref:dWPVP} 
B.\ de Wit, R.\ Philippe and A.\ Van Proeyen, 
\NP{B219,1983,143}.

\bibitem{ref:FGKVP} 
S.\ Ferrara, L.\ Girardello, T.\ Kugo and A.\ Van Proeyen, 
\NP{B223,1983,191}.

\bibitem{ref:KTN} 
M.\ Kaku, P.~K.\ Townsend and P.\ van Niewenhuizen,
\PR{D17,1978,3179}. \\
M.\ Kaku and P.~K.\ Townsend,
\PL{76B,1978,54}.\\
M.\ Kaku, P.~K.\ Townsend and P.\ van Niewenhuizen,
\PR{D19,1979,3166}, 3592.

\bibitem{ref:vN} 
P.\ van Niewenhuizen,
Phys.~Rep. \andvol{68,1981,189}.
\end{thebibliography}
\end{document}